\documentclass[sigconf,authorversion,nonacm,table]{acmart}
\PassOptionsToPackage{table,xcdraw}{xcolor}
\usepackage{algorithm}
\usepackage{algpseudocode}
\usepackage{framed,multirow}
\usepackage{makecell}
\usepackage{tabularx}
\newcounter{magicrownumbers}
\usepackage{cleveref}
\usepackage{pdfpages}
\usepackage{hyperref}
\newcommand\rownumber{\stepcounter{magicrownumbers}\arabic{magicrownumbers}}

\AtBeginDocument{%
  \providecommand\BibTeX{{%
    \normalfont B\kern-0.5em{\scshape i\kern-0.25em b}\kern-0.8em\TeX}}}

\usepackage{adjustbox}
\usepackage{fancyhdr}
\begin{document}

\title{Retrieval Augmented Verification for Zero-Shot Detection of Multimodal Disinformation}

\author{
Arka Ujjal Dey, Artemis Llabrés, Ernest Valveny and  Dimosthenis Karatzas
\\
Computer Vision Center, \\Universitat Aut\`{o}noma de Barcelona, Spain \\
}

\begin{abstract}
The rise of disinformation on social media, especially through the strategic manipulation or repurposing of images, paired with provocative text, presents a complex challenge for traditional fact-checking methods. In this paper, we introduce a novel zero-shot approach to identify and interpret such multimodal disinformation, leveraging real-time evidence from credible sources. Our framework goes beyond simple true-or-false classifications by analyzing both the textual and visual components of social media claims in a structured, interpretable manner. By constructing a graph-based representation of entities and relationships within the claim, combined with pretrained visual features, our system automatically retrieves and matches external evidence to identify inconsistencies. Unlike traditional models dependent on labeled datasets, our method empowers users with transparency, illuminating exactly which aspects of the claim hold up to scrutiny and which do not. Our framework achieves competitive performance with state-of-the-art methods while offering enhanced explainability.
\end{abstract}
\maketitle

\section{Introduction}
\begin{figure}
\begin{center}
\includegraphics[width=.5\textwidth]{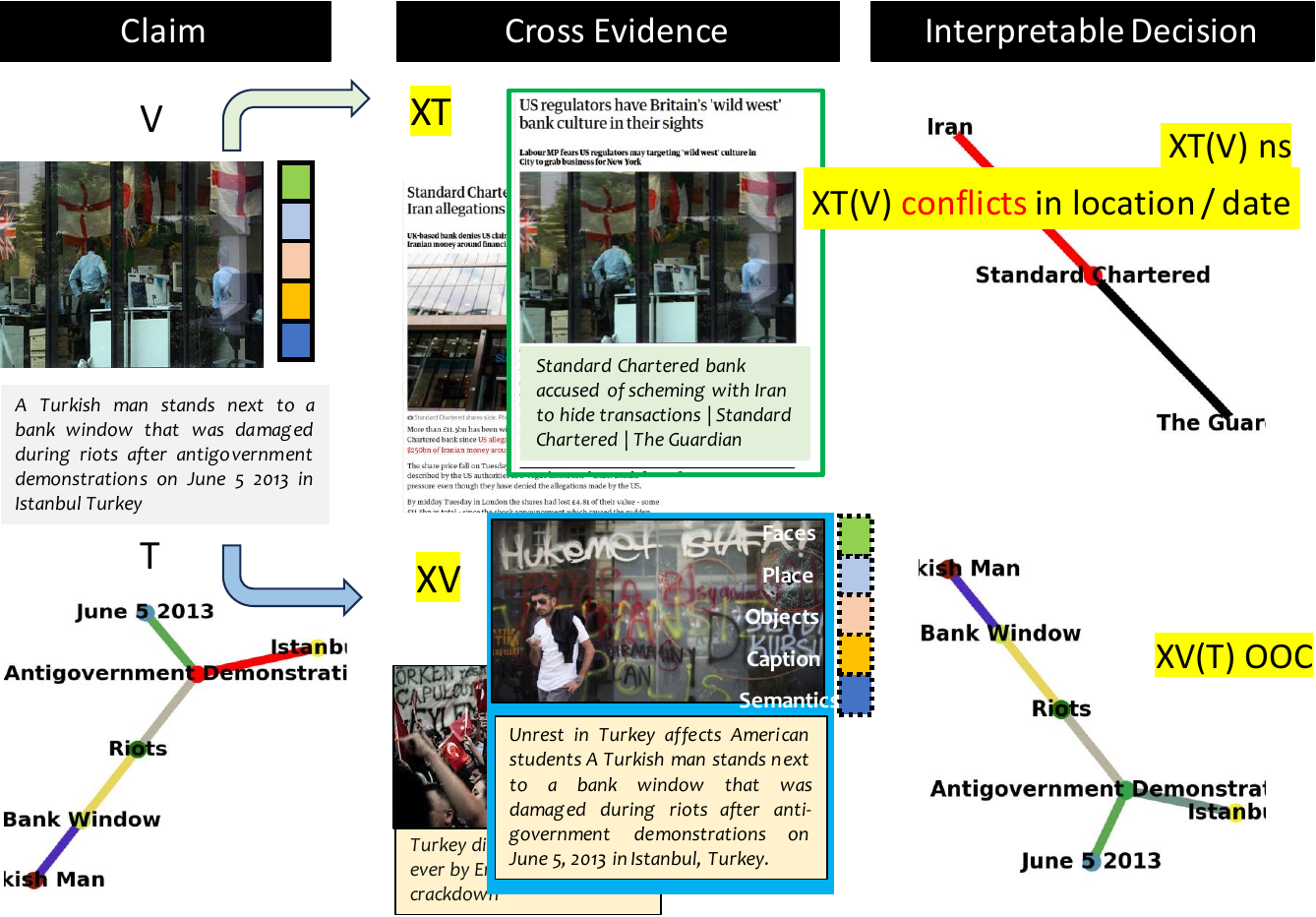}
\caption{Core Idea: Fact-checking Social Media Posts against News Websites: Example of an image-text claim where the image has been used out-of-context. XV(T) is Visual cross-evidence from text claim T, and XT(V) is Text cross-evidence from image claim V.  We show through a graph-based text representation that `riots' and `June 5 2013', among others, are supported by the XV(T)  through similarly colored nodes and edges. However, the retrieved image is not visually similar to the original one, and thus, this is judged as Out-of-Context (OOC). Furthermore, XT(V) does not support T in terms of matched nodes or edges  but instead conflicts regarding the location context (`Turkey' vs. `Iran').}\label{fig:mm}
\end{center}
\end{figure}
\textbf{Disinformation through Social Media} posts can be characterized as retelling an existing story, albeit with an ulterior motive, usually with a \textbf{visual aid}. This form of Multimodal disinformation can be particularly dangerous because images are a powerful tool for propaganda, often evoking deep emotions. The inclusion of images in Social Media posts \cite{li2020picture} leads to increased engagement through likes and shares, thus amplifying the perceived credibility of the accompanying text, even when the content is false. This engagement and amplification accelerates the spread of  disinformation for multimodal Posts\cite{hameleers2020picture}. The widespread availability of easy-to-use image manipulation tools exacerbates the issue, allowing users with little to no technical expertise to re-purpose or alter visuals. As a result, manipulated or out-of-context images have become a major driver of disinformation, making it increasingly difficult to discern fact from fiction in multimodal posts.

The kinds of manipulations typically seen in images include manipulation of the scene text (changing text on propaganda posters), changing attributes of visual elements (making faces smile, or swapping faces). Often, the image used is real but has altered text. These text manipulations might use the visual cue to spread lies about an unrelated event or add or skip details from the text for a particular motive. There are multiple such ways to manipulate the original claims. By design, these claims are without editorial oversight and accessible to a vast population who otherwise may not have access to multiple information sources. This implies the need to fact-check these posts and clearly explain which parts of the posts are fake.

Manual fact-checking  in \textbf{Traditional Journalist approach} involves conducting interviews, asking questions,  cross-referencing multiple credible sources, and seeking out reliable evidence. This process is thorough, aiming not just to determine whether a piece of information is true or false, but also to identify which specific parts are accurate or misleading. Media houses or individual journalists take responsibility for the information they disseminate.

But manual fact-checking is time-consuming, and thus in practice, fact-checking efforts have mostly focused on viral content, given its wide reach and potential impact. However, even then, the verification process often lags, allowing the viral content to cause damage before it can be corrected \cite{cooke2017posttruth}. Additionally, non-viral posts, which frequently go unchecked, can be equally or even more problematic as they continue to spread false narratives unverified.

\textbf{Automated Tools } aimed at detecting such manipulations in isolation requires labeled datasets, which are scarce, leading to the development of tools to create synthetic datasets. Previous efforts at creating synthetic datasets focused on swapping captions \cite{jaiswal2017multimedia} or replacing entities \cite{sabir2018deep} and generating pseudo-fakes. In \cite{luo-etal-2021-newsclippings}, we see the use of a language vision system to create synthetic fakes by mapping news clip images to semantically similar but unrelated (in reality) captions. The authors observed that machine-driven image re-purposing is now a realistic threat and provided samples that represent challenging instances of mismatch between text and image in news that can mislead humans. This was followed up in \cite{shao2024detecting}, where further automated manipulations were introduced through complete swaps or slight attribute changes, creating a further realistic out-of-context dataset. The creation and accessibility of these tools for creation of synthetic datasets actually facilitate the large production of fake news. The fake image-text pairs thus generated are so convincing that they could no longer be judged in isolation but only through support from external knowledge, almost like how journalists fact-check news by looking for supporting or contesting evidence.

In \cite{abdelnabi2022open}, we see our inspiring idea, where the authors use \textbf{automatically web-scraped external evidence} to detect out-of-context (OOC) usage of image-text. The authors introduce the idea of  \textbf{cross evidence} to fact-check multi-modal posts or claims, where the image claim is reverse searched to find \textbf{text cross evidence}, and text claim is searched to find \textbf{image cross evidence}. This represents our core idea of finding cross evidence to support or reject multimodal claims. However, the underlying issues persisting are with the \textbf{retrieval of relevant evidence}, inherent \textbf{bias} in such systems due to training data and the \textbf{lack of explanation} in the final judgment. 

The style of the text (claim) biases the results (evidence) it returns, meaning, when a story is retold subjectively in social media, it loses the style of the source (news website), affecting the retrieval of relevant evidence from direct searches. Further, because of the supervised learning-based setup, this detection of fake news is reduced to a binary classification problem, neglecting all intermediate stages. We believe that it is not enough to say something is fake or verified; the system must be able to explain such judgment. We argue that supervised learning-based end-to-end fact-checking systems are susceptible to biases not just in the final decision but also in the upstream selection of evidence that leads to the decision. Also these claims often involve recent events on which systems trained on historical data are prone to fail. Finally, such detection in isolation tells us nothing about the provenance of the claim image. Verifying the authenticity of such claims requires more than just a surface-level inspection; it necessitates cross-referencing with reliable sources and looking for supporting or contesting evidence. This approach is akin to journalistic fact-checking, where the goal is to substantiate the claim by investigating its origins and corroborating details. Only through such diligent verification can we hope to counter the spread of disinformation effectively.

In this work, we rely on the idea of using external evidence but try to overcome the main limitations of previous approaches, which can be summarized in the following points:
\begin{itemize}
\item Fact-checking relies on the \textbf{retrieval} of good evidence, which is often affected adversely by claim visual quality and text style.
\item Learning-based approaches need \textbf{labeled data}, which is difficult to obtain, and synthetic datasets often don't capture the distribution of actual fake news.
\item \textbf{Black box} binary classification often renders the final output opaque. This is particularly relevant for fact-checking, where explaining is often as crucial as prediction.
\end{itemize}

This leads to the following philosophy where instead of a binary classification about authenticity, we highlight which parts of a social media claim differ or agree with external evidence. We position this not as a tool to detect fake news automatically, but to aid and enable human users find relevant evidence and  while clearly point out the similarity and differences with the claim. Our core idea (Figure \ref{fig:mm}) is to represent the text and image in a way such that they can be easily compared through rule-based matching for support or conflict with evidence. For the text, we represent it as an \textbf{Entity Relationship graph}, while for the image, we use \textbf{Pretrained Language Vision} systems to represent it in terms of the objects, faces, places, scene text in the image. With such \textbf{structured representations}, we can not only say if two texts or two images are similar, but we can explicitly state what they agree with and what they conflict about leading to \textbf{improved interpretability}. Our \textbf{zero-shot rule-based matching} is an alternative to data-driven learning of fake versus real claims. This zero-shot approach with structured representation  allows us to place a social media post in the context of news articles clearly highlighting the supporting and conflicting elements, and, therefore, enriches the final decision with explainability. To address the challenge of evidence retrieval resulting from variations in text style, such as verbosity or subjective retelling, we present a feedback-based retrieval method that utilizes entity relationship graphs to iteratively refine the search results.

The visual aid usually consists of some existing image being repurposed or manipulated; thus, finding its real-world origin or provenance would unmask the truth. However, the danger of increasing realistic deepfakes means we should look only at credible sources. Given that the purpose is misuse, the most harmful choices of images for re-purposing are those that are already rich with visual information rather than generic or symbolic. These are also the kinds of images that are often credibly reported in news media. Thus, we source evidence only from news websites.

In summary, our main contributions are:
\begin{itemize}
\item We propose a Structured \textbf{Representation} of Text in terms of Large Language Model \textbf{LLM}-aided Entity Relationship graphs and Pretrained Visual Features that allow us to do rule-based matching against data-driven learning for Out-of-context Detection.
\item We propose a Feedback-based \textbf{Retrieval} that iteratively improves search results by leveraging our structured representation and exploring unmatched nodes. This retrieval is aided by an LLM.
\item We propose a \textbf{Zero-shot Verification and Explanation} scheme that applies strict matching rules to the structured representations of the claim and evidence to generate interpretable results.
\end{itemize}

\section{Related Works}

In this section, we look at the related work in terms of 1) fact-finding or evidence retrieval strategies, 2) the supervision used to learn, 3) the explanations, if any provided, as part of the reasoning, and finally, 4) the bias in the designs of the task and dataset. 

Several works study the detection of multi-modal misinformation \cite{abdelnabi2022open,aneja2021cosmos,luo-etal-2021-newsclippings,khattar2019mvae,jin2017multimodal}. Some of them deal with a small scale human-generated multi-modal fake news \cite{khattar2019mvae,jin2017multimodal}, while others address out-of-context misinformation where a real image can be paired with a swapped real text often with manipulated textual and location data as in \cite{sabir2018deep} or even without any manipulation \cite{abdelnabi2022open,aneja2021cosmos,luo-etal-2021-newsclippings}.

\subsection{Retrieval Based Reasoning  }
The use of external evidence has been explored in Vision Language tasks, but mostly related to Visual Question Answering   ~\cite{Marino2019OKVQAAV, Li2020BoostingVQ,wu2016ask} leveraging publicly available external knowledge bases. In the case of VQA, primarily it is the question words~\cite{Marino2019OKVQAAV} along with visual cues in the form of scene labels~\cite{Marino2019OKVQAAV}, detected entities~\cite{Li2020BoostingVQ}, or predicted visual attributes~\cite{wu2016ask}, that are used to retrieve knowledge from external sources. Once retrieved, the knowledge facts are incorporated into the \mbox{\emph{answer generation}}. Success in the VQA setup has led to similar architectures and labeled datasets being adopted in the misinformation detection task. In question answering, we gather evidence to answer a specific question about the input; in fact-checking, we look for evidence that verifies the claims. This verification usually entails the retrieval of related evidence, followed by binary classification or threshold-based similarity checks. 
In Factify \cite{mishra2022factify}, Mocheg\cite{yao2022end}, CCN \cite{abdelnabi2022open}  we find examples of multi-modal fact-checking based on knowledge. While Factify is more of a reasoning task with only one piece of evidence for each modality,  Mocheg uses web-scraped image text evidence to verify text-only claims. However, this retrieval is unrestricted, leading to the possibility of retrieving falsified evidence from a propaganda medium, corrupting the final decision. 

In \cite{abdelnabi2022open}, we encounter a framework for verifying image-text claims with multiple multi-modal evidence, which are retrieved from news websites. This news website-based retrieval adds credibility to the evidence source. The text and image claim parts are queried to generate cross-image evidence and cross-text evidence. The motivation behind cross-evidence is to check if the image or the text has been used in a similar context. This cross-modality search also means the final retrieved evidence is in the same modality as the claims, ensuring easy uni-modal comparison. The authors propose a memory network-based binary classifier. While the memory network is responsible for the relevant evidence collection, the model is not explicitly designed to point out the clenching evidence (maximally relevant single evidence) that leads to the decision. \cite{papadopoulos2023red} addressed this issue with their focus on identifying the relevant evidence first. 

\subsection{Interpretable models} While \cite{papadopoulos2023red} allows for explicitly pointing out which evidence led to the decision, it can not give fine-grained information expressing which parts of the claim are supported by this evidence. This is often a key requirement in Fact-checking, where the reader is interested in knowing how exactly the evidence supports or contests the claim. In \cite{shao2024detecting}, the authors propose a supervised multi-label classification scheme trained to add a degree of explainability to their model. The supervised multi-label classifier detects complete changes or swaps in the image or text regions based on its training on a synthetically augmented version of the dataset proposed in \cite{abdelnabi2022open}. While labels render the model interpretable in terms of the output, it is susceptible to bias due to the synthetic augmentation as well as the final supervision. We argue for structured representation-based reasoning, where we focus on generating representations that can be easily applied to rules and checked for consistency without any learning involved. Such an approach lends itself to be interpretable by design and free from training bias.

\subsection{Supervision is Bias}   
The idea of what is fake and what is not is inherently challenging. In ways, the focus is more about finding and emphasizing differences and not just similarities. The human understanding of what is fake is often a result of complex rationale, prior experience, and the latest evidence. It results in arguments and debates that explain the fairness of a post in terms of support and conflict with external independent evidence and sources. It is more akin to the task of fact-finding and applying a set of rules, where the fact-finding is often the bulk of the effort, and the rules are clear and unambiguous. Not unlike legal proceedings or judgments. In the supervised learning setup, this is, however, reduced to a binary label, neglecting all intermediate stages. Treating the problem as a data-driven, learnable task. These methods are, in general, binary classifiers trained with supervision, either from human annotations or from the sampling scheme used to generate them. Such a system has an inherent bias towards the dataset it has been trained on with poor generalization. Even when external knowledge is incorporated into the pipeline to add generalization, the choice of external knowledge is guided by the final supervision.

\begin{table}
\setcounter{magicrownumbers}{0} 
\small
 \caption{Statisics on Data Splits .
  samples: Total number of posts in the dataset; \# Multi-modal: Number of post containing images; \#XT: Number of selected samples with external textual evidence; \#XV: Number of selected samples with external visual evidence; \#S: Total Number of samples, searched for external evidence. }
 \label{tab:dataset_split}
 \setlength\tabcolsep{3 pt} 
  \begin{tabular}{|l|l|l|l|l|l|}\hline
    Acronym & Dataset              &  samples                      & \#Multi-modal       & \#XT / \#XV  &  \#S  \\ \hline
    B        &bcn19             &  300943                       & 168387            & 40 / 154  &  154 \\
    M1       &mena\_aggr        &  3074                         & 1641              & 42 / 42   &  274 \\
    M2       &mena\_ajud        &  1799                         & 953               & 48 / 48   &  204\\
    O        &openarms          &  7123                         & 1140              & 20 / 20   &  230 \\ \hline
    N        &NewsCLIPIngs &  7233                         & 7233              & 5278/7233 & 7233    \\ \hline
  \end{tabular}
 \end{table}

\begin{figure}
\begin{center}
\includegraphics[width=.49\textwidth]{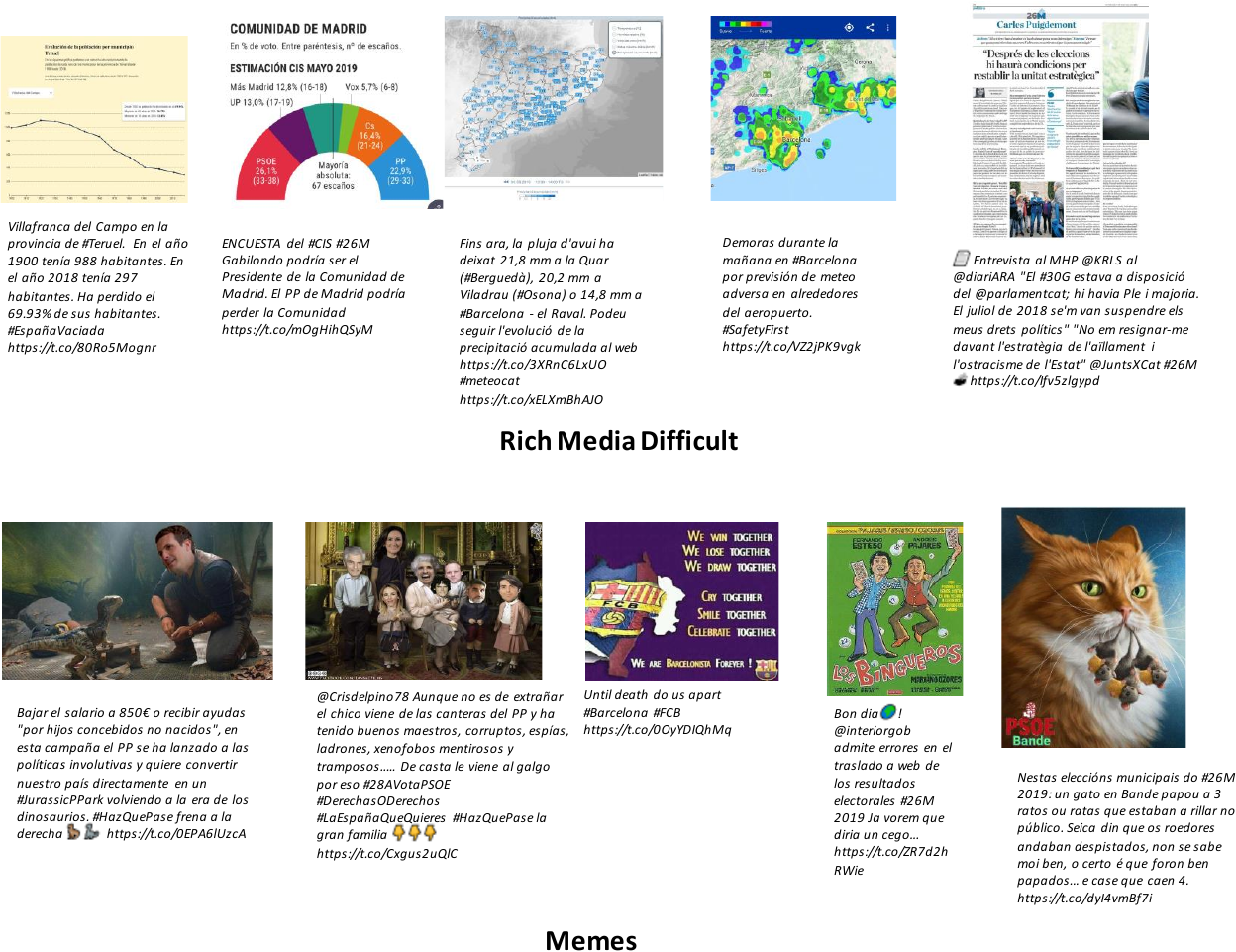}
\caption{ Samples for Verification (unsuited for our method): Rich Media Content and Memes }\label{fig:meme}
\end{center}
\end{figure}
\begin{figure}
\begin{center}
\includegraphics[width=.49\textwidth]{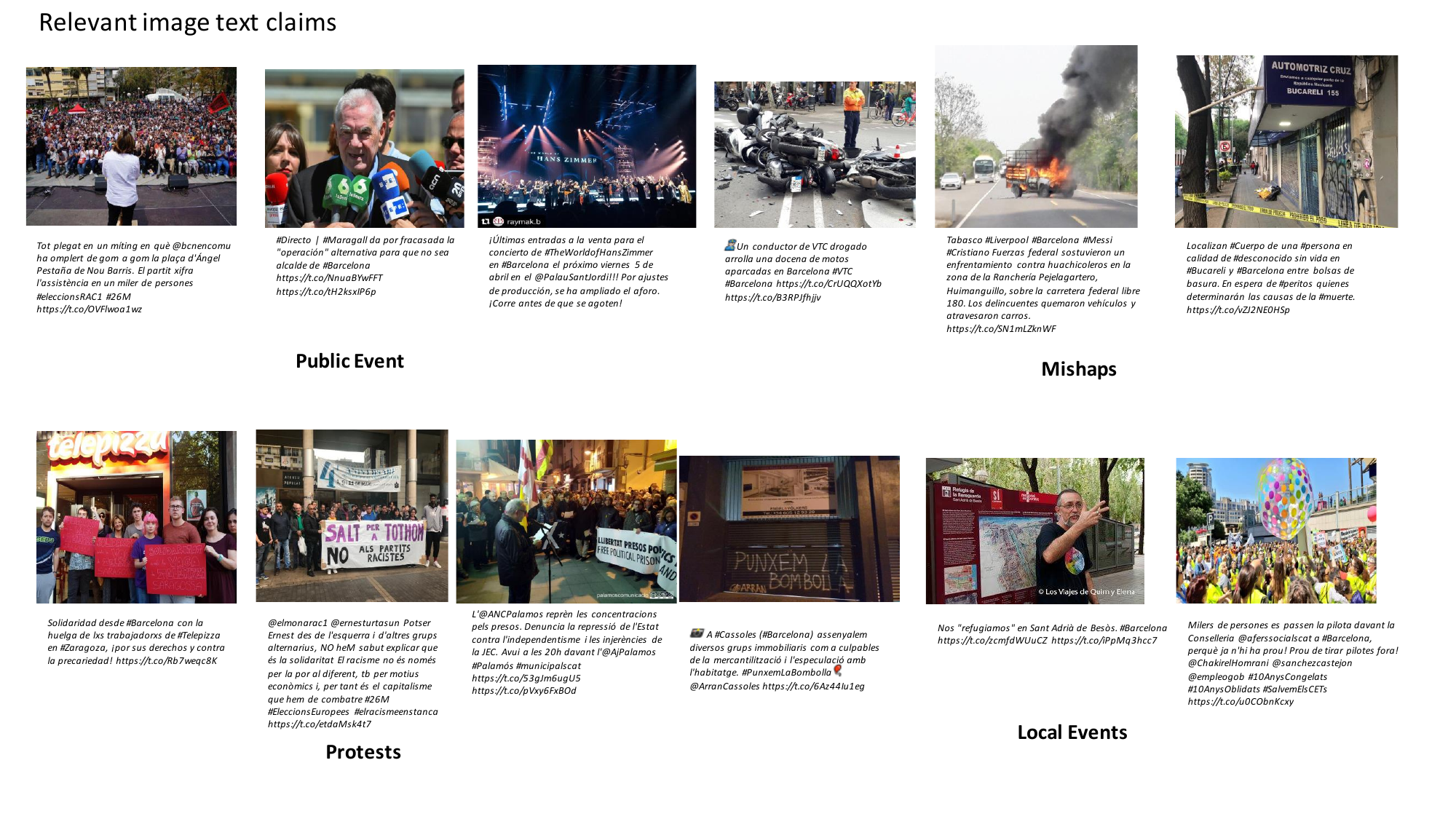}
\caption{Samples for Verification (Use case): Samples with natural images related to news event  are best suited for our pipeline. }\label{fig:acp}
\end{center}
\end{figure}
\begin{figure}
\begin{center}
\includegraphics[width=.48\textwidth]{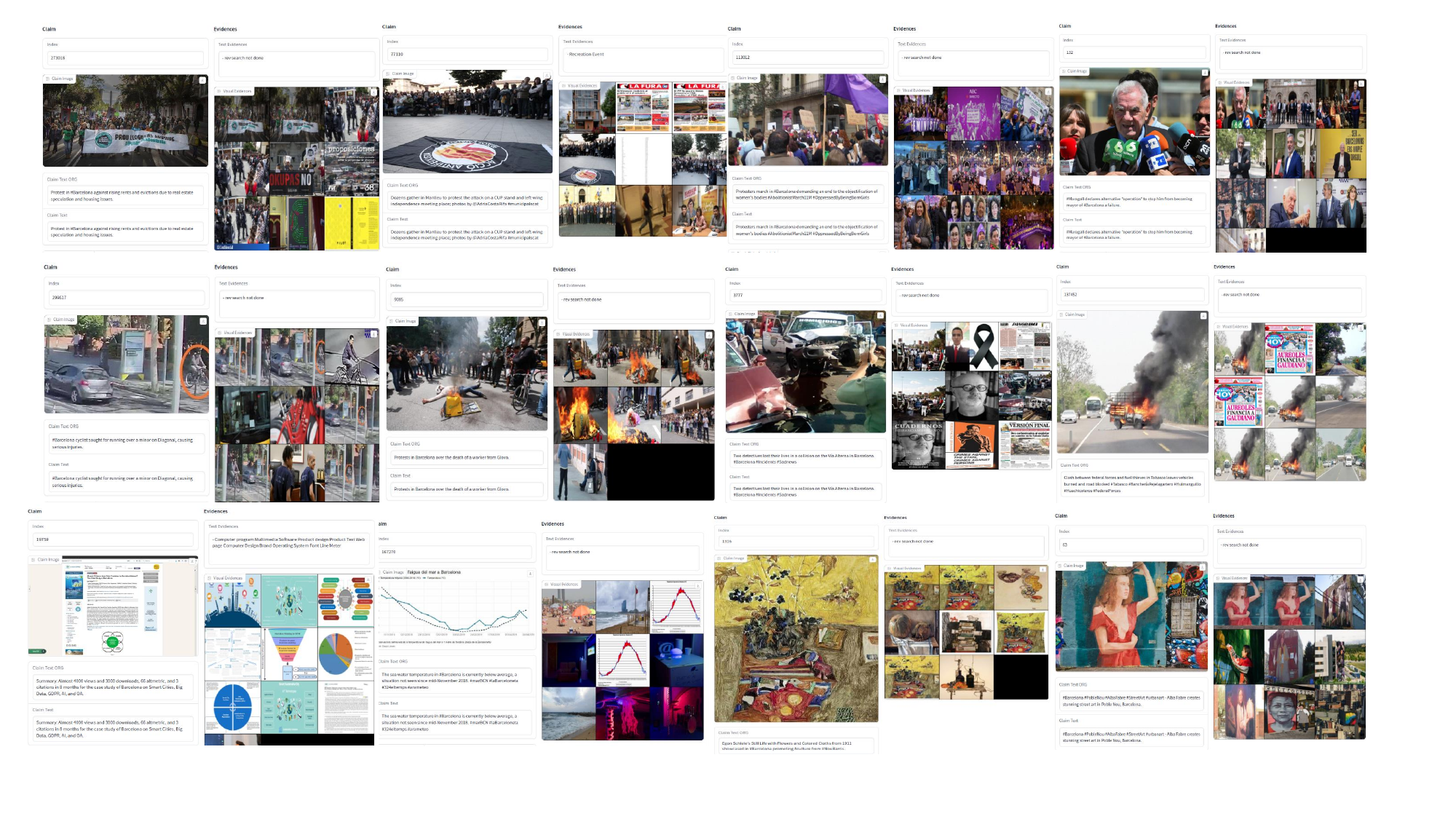}
\caption{Remiss : Samples which are annotated as Pristine or True News, based on retrieved evidence}\label{fig:prist}
\end{center}
\end{figure}

\begin{figure}
\begin{center}
\includegraphics[width=.49\textwidth]{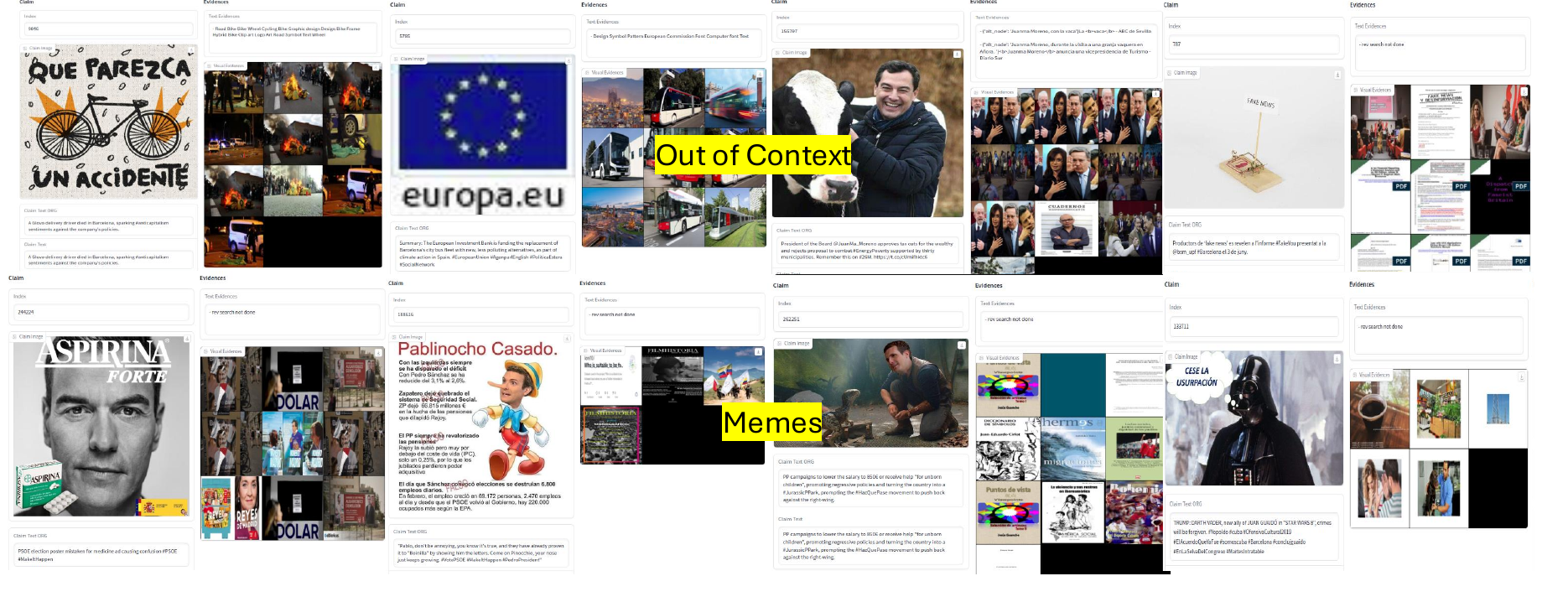}
\caption{Remiss : Samples which are annotated as Fake News, based on retrieved evidence}\label{fig:fake}
\end{center}
\end{figure}

\section{Datasets }
The datasets used in this study comprises unlabelled tweets, systematically gathered by a team of journalists over a period of time, with a focus on potential hateful or deceptive content. We use four different datasets organized around two main topics. The first one includes propaganda, hate speech, and false claims concerning elections (bcn19 or 'B') and the second topic is about immigration (mena\_aggr or `M1', mena\_ajud or `M2', openarms or `O' datasets). We refer to this dataset collectively as \textbf{Remiss}. Our investigation centers on multimodal instances where the textual tweet is accompanied by an image. 

Table \ref{tab:dataset_split} shows some statistics of the dataset. We can see (column \emph{Multi-modal}) that a substantial proportion of the tweets are associated with images. Nevertheless, these samples are not associated with any provided external evidence, which is critical for our evidence-based methodology. Consequently, we sampled some posts  (column \#S) and searched for external textual and visual evidence (\#XV and \#XT). For all datasets, we only consider samples for verification that have both text and visual evidence.

It's important to note that in real-world scenarios, obtaining evidence may not always be straightforward or complete, thus  we also wanted to study the effect due to lack of good evidence. Therefore, for dataset `B', we verify samples that have either text or visual evidence. Through our experiments, we will demonstrate that incorporating a \textbf{Human-in-the-Loop} approach can enhance automated evidence collection, leading to higher quality evidence and improved accuracy in the verification stage. This approach acknowledges the challenges of evidence availability and leverages human judgment to fill gaps and ensure the reliability of collected evidence.

Finally, we also compare our method against \textbf{NewsCLIPings} \cite{luo-etal-2021-newsclippings}, a public benchmark collected from news portals like The Guardian, BBC, USA Today, and The Washington Post . The fake examples in this dataset are swapped image caption pairs.  For NewsCLIPings we use the evidence XT and XV provided as part of the work in CCN \cite{abdelnabi2022open}. It is to be noted that the image text pairs used here were curated from News websites and thus the reverse search results exist for 73 \% of samples.

\subsection*{Limitations due to Alignment and Rich Media Content} Our approach to fact-checking multimodal posts hinges on the alignment of images and text conveying a shared concept. Explicit and concrete shared concepts are easier to verify, but symbolism in text or visuals complicates comparisons and evidence retrieval. Symbolic visuals not only make comparisons with text difficult but also hinder evidence gathering when using the text as a query. For instance, a tweet about a protest with a symbolic image is harder to verify compared to one with a real protest image, which can be cross-verified by locating its source. Our method is most effective when images depict specific details mentioned in the text and are semantically aligned, as illustrated in Fig. \ref{fig:acp}.

Additionally, our method struggles with rich media content like charts or diagrams. Fig.  \ref{fig:meme} showcases examples of such content in the top row, where text-heavy images with poor visual quality are challenging to compare or find online. We include these samples to demonstrate our method's handling of them. The bottom row shows memes, which are easily classified as fake due to the absence of supporting evidence, as depicted in Fig.\ref{fig:ret}.

\subsection*{Annotation: What is Fake} \label{whatisf} We annotate the collected claim samples as `Pristine' or `Fake' with the help of human annotators. This annotation process involves examining available evidence on the internet and applying human expert understanding. It is not limited to the evidence retrieved in the first step. However in the absence of inconclusive evidence the annotator is encouraged to mark the sample as fake. In other words, the annotator makes a decision on each sample based on their own comprehensive research, rather than relying solely on the initially retrieved evidence.

\paragraph{\textbf{Pristine News must have supporting Evidence}} We label samples as Pristine when we can validate the claim with external knowledge in News websites. Fig.\ref{fig:prist} we show some samples and their corresponding retrieved evidence.  
\paragraph{\textbf{Fake News has no Evidence, or has conflicting Evidence}}We label samples as fake, either due to lack of any evidence of it in news websites, or if the evidence conflicts with some specific detail, as illustrated in Fig.\ref{fig:fake}. In row 1 we show samples exhibiting Out-of-context usage. These are cases of symbolic images paired with news text (first two from left), or an old image repurposed with a different text (third from left). For these kind of fake news we find evidence image using the text that contests the image. In row 2, we see examples of Memes, which are trivially proved as fake due to the lack of any meaning evidence. While we see that some evidence were retrieved, they are rejected in the  checks.

\begin{figure}
\begin{center}
\includegraphics[width=.45\textwidth]{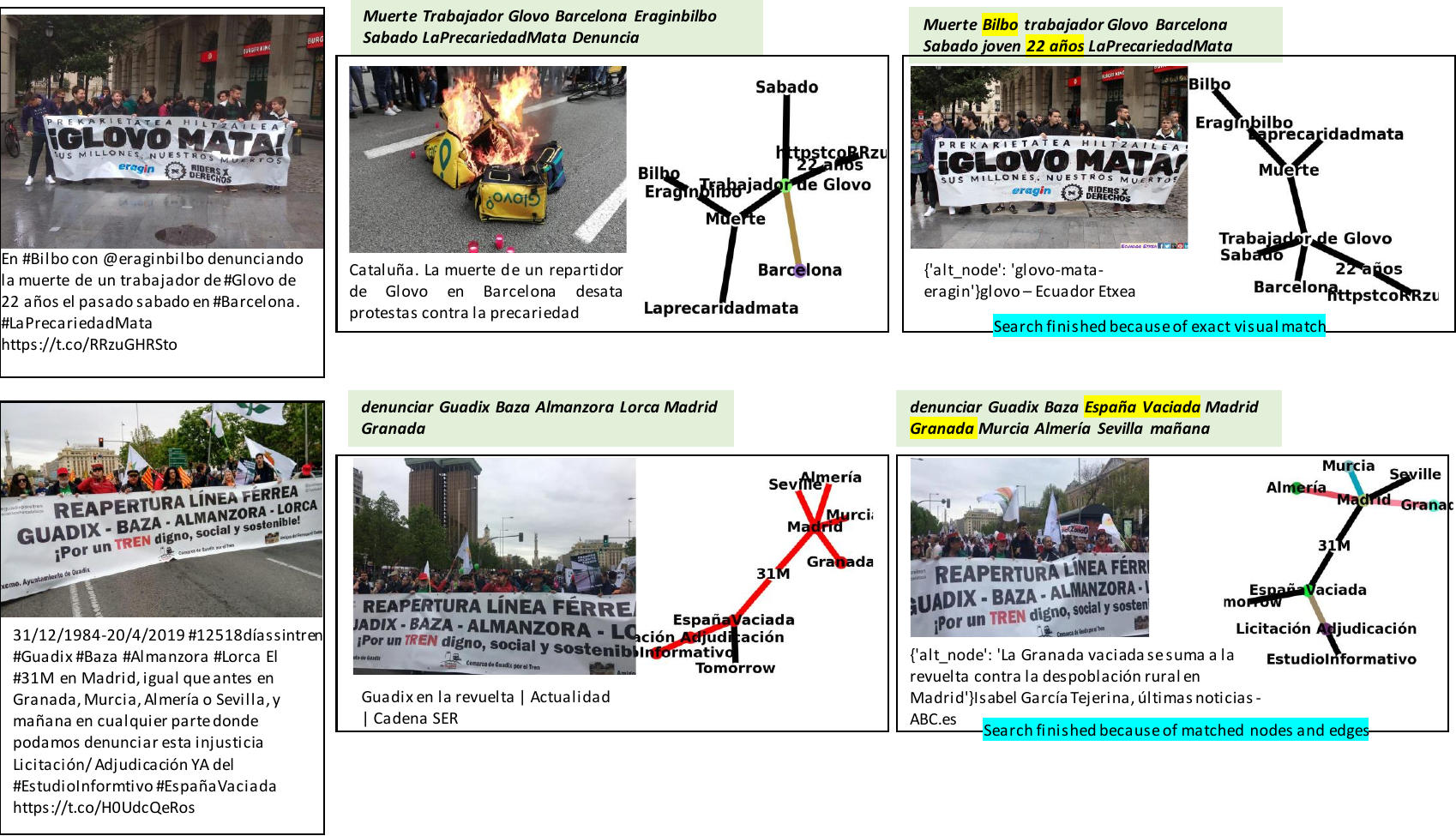}
\caption{Feedback Based Retrieval: Use unmatched nodes to construct text search term}\label{fig:fdr_remiss}
\end{center}
\end{figure}

\section{Method : Retrieval Augmented Verification}

Our primary assumption about the claim is that it is presented in the format of a Text-Image pair. Given the claim couplet, we are tasked with verifying the content with evidence. The first part of our work entails retrieving external cross-evidence based on Internet searches. These retrieved evidence are then ranked based on similarity to identify the relevant evidence. Our key insights are 1) fine-grained structured representation of the claim and evidence, which allows us to explicitly point out supports and conflicts while also being interpretable, and 2) supervised learning leads to bias, and thus, a zero-shot approach to detect the said conflicts and supports is more desirable. In Fig. \ref{fig:main} we present our framework, based on the following components.

\begin{itemize}
    \item Multi-modal \textbf{Feedback based Evidence Retrieval} guided by Entity Relationship (\textbf{ER}) graphs
    \item \textbf{Structured  Representation} of text as Entity Relationship Graphs and images in terms of pretrained visual features 
    \item \textbf{Comparison Metrics} of Entity Relationship Graph (Graph Match)  and images (Image Match) leading to 
    \textbf{Interpretable}  \textbf{Verification} of Claim with Evidence  in terms of Supports and Conflicts
\end{itemize}

As we can see in Fig. \ref{fig:main}, the claim image is represented with a set of visual features while the claim text is converted into an Entity Relationship Graph. The image is used through reverse search to find cross-textual evidence, and the text is used to find cross-visual evidence. The graph-based representation of textual evidence and the visual representation of visual evidence are matched against the original text and image to find supporting and conflicting evidence. The result of the matching is also used to refine the retrieval of cross-evidence. Finally, we can get an interpretable decision in terms of matched nodes, edges and visual features. 
\subsection{ Evidence Retrieval}
Given an Image-Text Claim, we define \textbf{Text Cross Evidence (XT)} as the text evidence obtained by reverse search with the Image Claim, and similarly, \textbf{Visual Cross Evidence (XV)} as the visual evidence obtained by direct search with the Text Claim. In Fig.  \ref{fig:main}, we show an example of a text-image claim and the retrieved text and visual cross-evidences. We collect evidence following the scheme below : 
\subsubsection*{ \textbf{Visual Evidence}}
We query the \textit{Google} powered \textit{Programmable Search Engine}  with the text claim to collect Visual Cross Evidences \textbf{(XV)}. When the text claim is brief and factual, it can be used in this manner to query the Internet to find close matches. We attempted finding Visual Evidence by directly searching with the text claim as query. However for real world fake news from social media posts retrieval was a challenge due to the subjective retelling of the story in the claim which is often different in style from it is reported in news websites. Further for verbose text claims, we often need to summarize the text and create specific search terms, using the structured representation of the claim. However, the specific search terms that might work for an image depends on the annotation provided in the website. Thus, we adopt a feedback-based image retrieval, where we use text and visual feedback from the retrieved Visual Cross Evidence \textbf{XV} and its contextual text \textbf{XVT} to guide and refine the search term generation.

\paragraph{\textbf{Feedback-based retrieval}}
We propose a feedback-based retrieval scheme that leverages our structured representation of the query (introduced in section Sec. \ref{sec:st_rep}) text to guide the search. In Fig. \ref{fig:fdr_remiss}, we illustrate how we can enhance initial retrievals by identifying unmatched nodes and adjusting the search query accordingly. We employ pretrained visual networks, detailed in Sec.\ref{sec:st_rep}, to score the similarities between the claim image \textbf{V} and retrieved evidence image \textbf{XV}, forming our visual feedback. This is combined with text feedback from comparing the graph representations of the claim text \textbf{T} and the contextual text scraped alongside the retrieved image  \textbf{XVT}. This combined visual and text feedback is used to propose a modified search term. Specifically, we obtain visual similarity scores for objects, semantics, place, face, and caption, which are communicated to the Large Language Model to refine the search string based on named entities related to semantics, place, and person from the graph.

As seen in Tab \ref{tab:dataset_split}, in this way we are able to collect visual evidence for 264 of the total of 862 sampled claims. However, this automated approach doesn't always succeed in gathering relevant evidence. Issues arise from overly complex search terms or irrelevant search results. Furthermore, the system heavily relies on similarity measures to determine evidence relevance. Our study demonstrates that involving a \textbf{Human-in-the-loop} for the evidence collection process, capable of adjusting search terms and evaluating evidence quality, can significantly improve the quality of the collected evidence.

\begin{figure}
\begin{center}
\includegraphics[width=.49\textwidth]{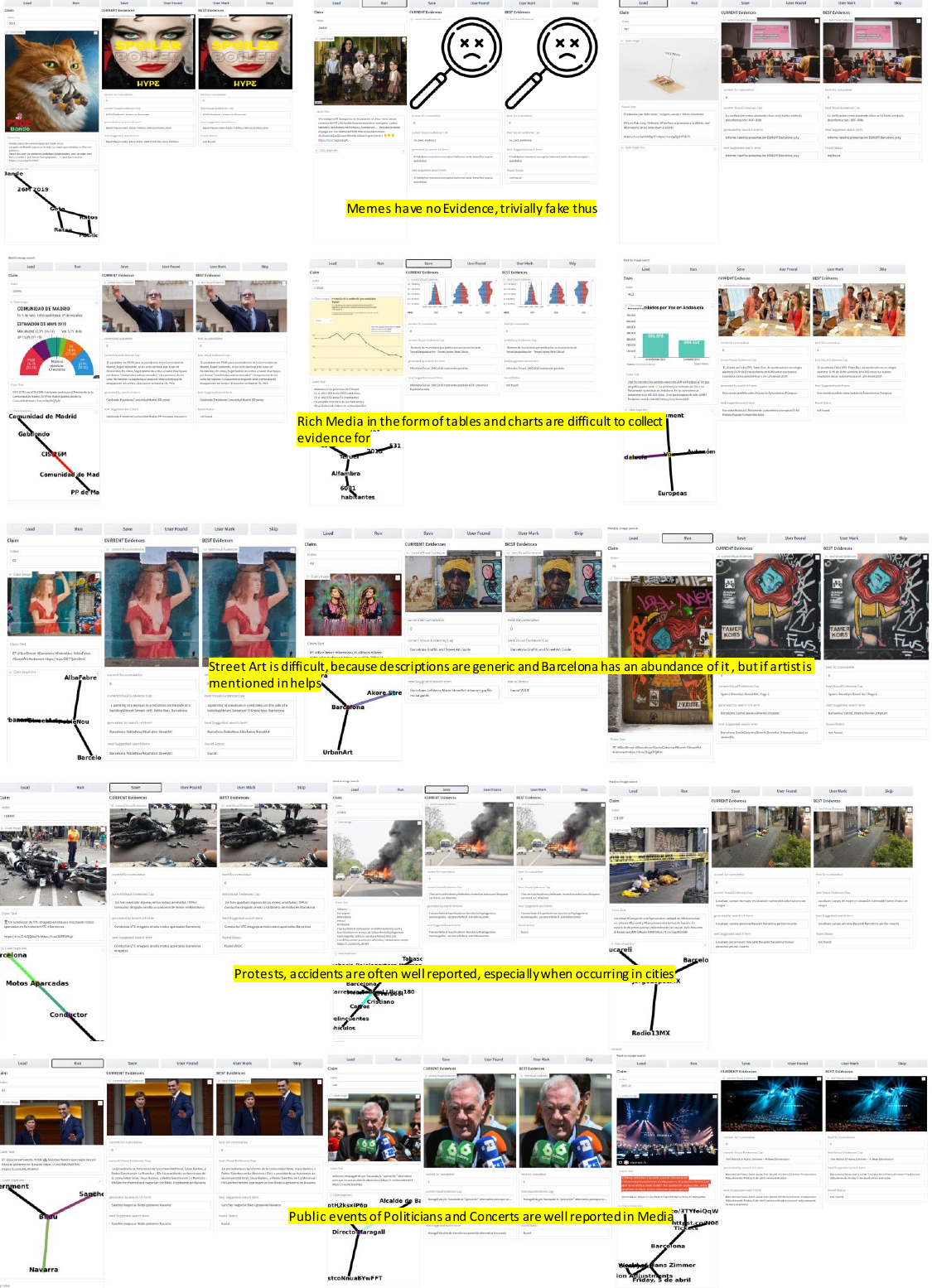}
\caption{Evidence Retrieval for Remiss: Across Post Types}\label{fig:ret}
\end{center}
\end{figure}

In Fig. \ref{fig:ret} we demonstrate our visual evidence retrieval across different types of multimodal posts. In row 1, we see that memes are trivially marked as fake, because of lack of supporting evidence. For rich media in row 2, we see that despite retrieving relevant evidence, the visual nature of graphs, charts makes it hard to compare with the collected evidence, thus these are posts our method fails to deal with. Row 3 shows examples of art, street art in this case, which is often hard to find evidence for without artist information. Finally in row 4, 5 we show that protests and public events are well covered in News Media, leading to correct evidence retrieval and verification.  The following \textbf{News Sources} were used for retrieving visual evidences:
\begin{itemize}
    \item \textbf{Remiss}: elpais.com, elmundo.es, abc.es, lavanguardia.com, larazon.es, naciodigital.cat, marca.com, granadahoy.com, ecuadoretxea.org, eldiario.es, diariocordoba.com, publico.es, beteve.cat, radiosabadell.com, elespanol.com
    \item \textbf{NewsCLIPings}: nytimes.com, irishtimes.com, stripes.com, hollywoodreporter.com, news.sky.com, justapinch.com, telegraph.co.uk, independent.ie, newyorker.com, cnn.com, washingtonpost.com, statnews.com, bbc.com, ibtimes.co.in, timesofisrael.com, dailymail.co.uk, nationalpost.com
\end{itemize}

\subsubsection*{ \textbf{Text Evidence }} We use \textit{Google Reverse Search} on the claim image to find Text Cross Evidence\textbf{ (XT)}. This Text allows us to find the context in which the claim image has been used in. However, this reverse search works best for content that is already widely published and is not as effective when tested on user-posted images on social media. We use the Google Vision API for our reverse searches. Google returns \textit{Complete} match result, \textit{Partial} and  also detected \textit{Visual entities}. The visual entities were often incorrect, and using them to prompt the language model leads to hallucinations. Thus we limit ourselves only to the  \textit{Complete} matches and reject the partial matches and the visual entities.

\begin{figure*}
\begin{center}
\includegraphics[width=.95\textwidth]{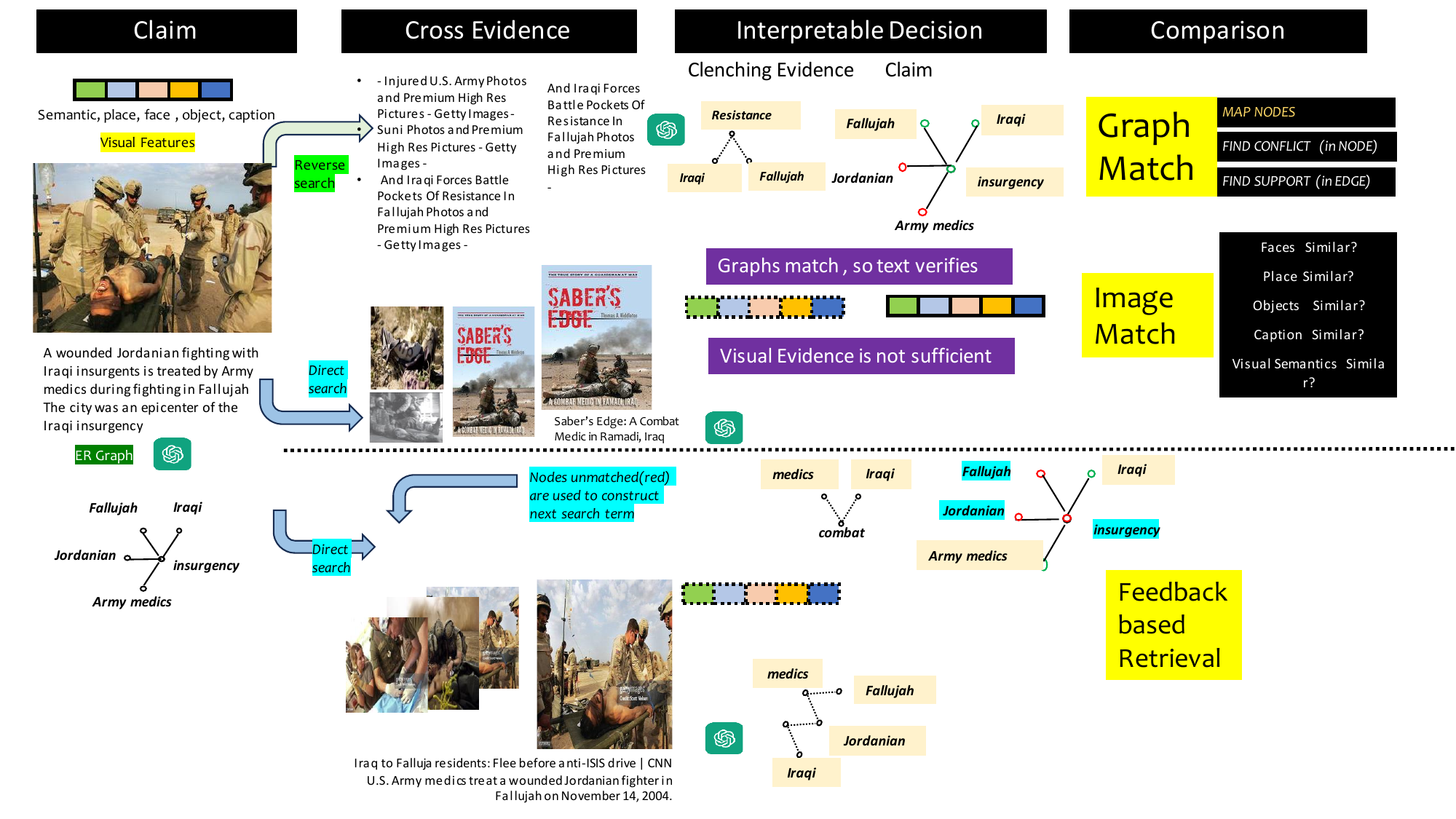}
\caption{RAV: Entails verifying Claims using Retrieved Evidence. However, instead of end-to-end  supervised, trained systems, we propose a zero-shot approach that uses structured representations for both verification and evidence retrieval }\label{fig:main}
\end{center}
\end{figure*}
 
\subsection{ Structured Representation}\label{sec:st_rep}

The images in the post are usually rich in famous personalities and landmarks, in addition to generic objects. The text, on the other hand, usually discusses named entities. This determines our feature choices detailed below. For images, we rely on detecting objects, faces and places, in addition to a generic semantic representation of the image. For the text, we use a graph-based representation that links entities represented in the text. 

\subsubsection{ Structured Representation for \textbf{Visual content}}\label{sec:visfeat}

The visual representation is obtained using standard pre-trained networks to extract the relevant visual information of the image: objects, faces, places and global semantics.

\paragraph{Objects} We use a pretrained detection model\cite{wu2019detectron2} to detect $N_o$ object bounding boxes, which are then encoded through a pre-trained Mask-RCNN Model\cite{he2017mask}. Similar to Cosmos\cite{aneja2021cosmos}, our visual encoding consists of RoIAlign and average pooling to generate visual object embedding $\{\mathbf{v^{obj}_{i}}\}$   $\in$ $R^{2048}$ (where \textit{i} = 1, . . , \textit{$N_o$}).

\paragraph{Faces} News images are often rich in personality faces, so we use a pretrained face detector \cite{mtcnnzhang2016joint} to detect $N_f$  faces, which are encoded through the pretrained facing embeddings\cite{ct_schroff2015facenet}  to generate visual face embedding $\{\mathbf{v^{face}_{i}}\}$   $\in$ $R^{512}$ (where \textit{i} = 1, . . , \textit{$N_f$}).

\paragraph{Place} Locations, or scene information is encoded through a pretrained network\cite{zhou2017places}, trained on  365 different types of places, to define our $\{\mathbf{v^{place}}\}$   $\in$ $R^{2048}$.

\paragraph{Semantic} We use a pretrained network \cite{vit_dosovitskiy2020image} to generate global image semantic features  $\{\mathbf{v^{sem}}\}$   $\in$ $R^{1000}$.

\paragraph{Caption} We use BLIP\citep{pmlr-v162-li22n} to generate an automated caption which we encode through BERT\cite{Devlin2019BERTPO}, to form  $\{\mathbf{v^{cap}}\}$   $\in$ $R^{768}$ 

\paragraph{Scene Text} Finally, we also use BLIP\citep{pmlr-v162-li22n} in a question answer mode to extract the scene text in the image from top left to bottom right. Which we encode through BERT\cite{Devlin2019BERTPO}, to form  $\{\mathbf{v^{sct}}\}$   $\in$ $R^{768}$ 

\paragraph{Final Visual Features}

\begin{equation}\label{feat:v}
\mathbf{v}= [ \mathbf{v^{obj}_{1,2,..,N_{o}}},   \mathbf{v^{face}_{1,2,..,N_{f}}}, \mathbf{v^{place}}, \mathbf{v^{sem}} , \mathbf{v^{cap}} ,\mathbf{v^{sct}} ]
\end{equation}
\noindent
\subsubsection{ \textbf{Build Graph:}Structured Representation for \textbf{Claim Text}}\label{sec:build_graph}

Given the plain text of a post, we want to represent them in a structured way in terms of the named entities and actions or relationships connecting them. Our principal idea is that comparing texts in terms of these graphs leads to a more fine-grained understanding of where the individual texts agree or conflict. We use a large language model to create this ER graph from plain text input. Our cautious use of LLMS is guided by detailed prompts, examples, and checks to ensure that we obtain a proper graph representation. We give the LLM specific instructions and examples about entity detection and relationship identification focusing on news stories, and we require a particular format that can be easily interpreted as a networkx graph \citep{SciPyProceedings_11}. This allows us to automatically check graph properties (connectivity, degree of nodes, walk, path)
leveraging the networkx library. We combine this with formatting checks and violations of our instructions to reject responses we deem unfit. 
 We define the nodes and edges as :
\begin{itemize}
 
   \item { \textbf{Nodes}} The named entities detected in the text by the LLM form the nodes. The named entity nodes are further enriched with facts from an external knowledge base. Similar to \cite{deykat}, we extract a set of candidate knowledge facts for each node and use the tweet text to select the most in-context candidate meaning according to semantic similarity. Thus, our Node representation consists of details about the type of entity and a contextually relevant description. We encode location and date entities in a specific hierarchical fashion, namely (city, state, country) and (day, date, month, and year), enabling exact correspondence and, thus, easy comparison.

   \item {\textbf{Edges}}  The edges connecting two named entities are defined using the LLM with an explicit extractive action and abstractive description. The action terms are restricted from being directly from the text, whereas their description is generated based on the LLM's knowledge about the action. This abstractive description allows us to map similar actions based on the description, like `protest' to `demonstration.'
\end{itemize}

\subsubsection{\textbf{Build Graph Conditional:} Structured Representation for \textbf{Evidence Text} } \label{bcond_gra}
The texts retrieved from evidences to be compared with the text in the claim are often widely different in their coverage of an event. While the claim may be a 280-character Twitter post, the web-scraped evidence text may be a few paragraphs. The graphs natively formed from varying lengths of text can have very different topologies, rendering them hard to compare. We represent an evidence text, focusing on the entities and relationships we have found in the claim text. 
We prompt the LLM to focus on the entities in the evidence text that are also present in the claim text graph and steer the detection around them. For the relationship we seek to validate, we pass the edges in terms of their participating nodes while masking out the detected actions in the claim graph task the LLM to predict them. This implies we force the evidence graph to have an edge between nodes if such nodes are also present in the claim graph and connected by an edge, thereby enforcing a similar graph topology.

\subsection{Comparison Metrics: Interpretable Verification with Graph Match and Image Match}

We interpret a News story as a collection of named entities that exist in the real world and a set of relationships or associations claimed by the news story. While in the past, people have tried to judge stories in their entirety as Fake or Pristine, we focus on trying to find out which parts of the story are true because we have evidence for it in the external world and which parts of the news can not be verified or are contested. Given an image text multi-modal post, our idea is to use the retrieved multi-modal evidence and independently verify the textual and visual aspects of the claim.

We compare the Claim Text and Image against the retrieved evidence, as shown in Fig \ref{fig:main}. Our multi-modal verification involves checking for 1) \textbf{Image match} by comparing the Claim Image and the Visual Cross Evidence  (XV) and 2) \textbf{Graph Match} by checking the Claim Text against the Text Cross Evidence (XT).
\subsubsection{  \textbf{ Image Match }} 
Visual verification with Image Match entails looking for exact matches from among the retrieved evidence.  We measure the images in terms of their pretrained feature similarity but only accept matches beyond high thresholds. Our \textbf{Image Match} scheme checks for visual aspects that are specific to news items encoding the images in terms of semantics, places, objects, faces, and automated caption. In particular,  we score the semantic, place, face, and object and automated caption features of the claim and retrieve evidence, in terms of the cosine distance of their embeddings. If any three of the 6 scores are similar beyond Image similarity threshold of 0.9 , we consider the images as matched..This allows us to explicitly point out how the evidence image matches or contests the claim image. 

\begin{figure}
\begin{center}
\includegraphics[width=.5\textwidth,keepaspectratio]{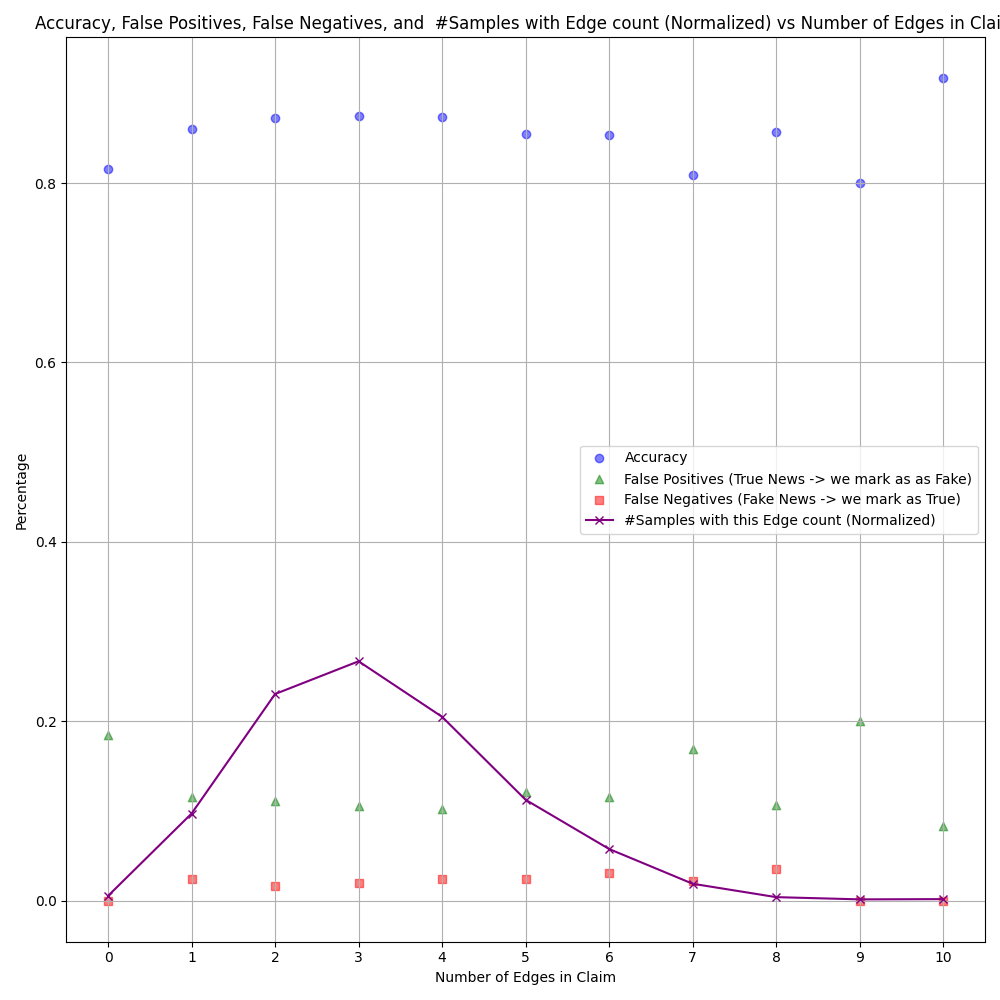} 
\caption{We plot the accuracy against the number of claims in the input text for the NewsCLIPpings dataset. Samples per edge count is the fraction of input samples with a particular edge count. We can see most claims consist of 2 to 4 edges.  }\label{fig:oocgraph}
\end{center}
\end{figure}

\subsubsection{\textbf{ Graph Match}} We measure the \textbf{Truthfulness of a text claim}, we compare the ER graphs from the Claim and Evidence text, in terms of \textbf{Supported Claim Edges} and \textbf{Conflicts in Nodes}  courtesy of our \textbf{Graph Match} scheme.  Our assumption is that for a True Claim,  every edge in the Claim graph must have a corresponding edge or walk in the Evidence graph. We check this through
\begin{itemize}
    \item Entity Matching :\textbf{ Map Nodes}  to check if similar nodes exist in the Evidence graph
    \item Conflicts in Nodes: \textbf{Find Conflict } to check if the said mapped nodes have any conflict in terms of location and date 
    \item and finally Edges Matching: \textbf{Find Support} to check if claim edges are connected by a semantically similar edge or walk in the evidence text.
\end{itemize}

\paragraph{ \textbf{Entity Matching}} 
Same entities may be represented slightly differently across texts, and thus, instead of an exact name-matching-based correspondence, we take into account the node details specific to the entity. Our prompting scheme enforces such details in an entity type-specific predefined format.
Thus, the nodes are encoded in terms of their name and description using pretrained word embeddings. We solve the node correspondence as a linear assignment problem through a modified Hungarian algorithm. The cosine distance between the node embeddings is used to define the cost matrix for the Hungarian. For a given node, we mask its cost related to the nodes of the same graph, forcing it to be mapped to nodes of the other graph.

\paragraph{ \textbf{Conflicts in the nodes} in terms of \textbf{Location and Date}}\label{ent_consistency}
While news articles may talk about similar people and events, their contextual details in terms of location and date distinguish them. Thus, in this step, we check if the nodes mapped in the previous stage are consistent in terms of the location and date type entities in their neighborhood. 
The hierarchical nature of formatting allows for the dealing of missing values in terms of city or state name or month. We deem a pair of matched nodes consistent in terms of location when they share the same location type entity in their neighborhood. A conflict is raised when the nodes have different location-type entities. A similar logic is applied to check for date checks. Any inconsistency in this stage leads to rejection of the mapping.

\paragraph{\textbf{Edge Matching} to find \textbf{Supported Claim edges} }
For a given edge (a,b, `<action>') in the claim graph, we check if the corresponding nodes $a^{'}$ and $b^{'}$ in the evidence graph, found using Hungarian in the previous stage, are connected in the evidence graph by a walk. However, the presence of a walk is not enough, as this walk could be contesting (disagreeing with) or verifying (agreeing with) the claim. Thus, we check the semantic similarity of this walk on the evidence graph against the claim edge. We collate the `<action>' terms along the walk and compare the cosine similarity against the claim `<action>' in terms of BERT embeddings \citep{Devlin2019BERTPO}. Thus, edges can be marked as `unconnected' if the nodes are not connected in the evidence graph by semantically similar `<action>' terms. Otherwise, they can be marked  `verified'.
This fine-grained marking of edges allows us to explicitly point out which parts of the claim were verified. This edge-matching scheme allows us to combine multiple evidence graphs and reason about the status of the claim edges simultaneously.

We output this fraction of claim edges verified as the measure of overall support for this claim given the evidence. This, of course, depends on the complexity and verbosity of the claim and the available evidence. In Fig \ref{fig:oocgraph}, we compare our accuracy against the number of claim edges in the sample. It demonstrates that claims that are either very generic (having 2 or less edges) or very verbose (more than 4 edges) are  where we struggle. For the majority of the samples the number of claims were around 3, for which our performance is comparable with the state-of-the-art. Finally we highlight against that our False Negatives are always lower than our False Positives, because as part of our design choice we wanted to prioritize detection of fakes over verification of Pristine samples.

\subsection{Parameters and Thresholds}
\begin{itemize}
    \item Entity or Node Similarity  Threshold is used to reject mapping during Node Matching using Hungarian. We set this value to 0.8, making sure only nodes that match beyond this threshold with their name and description field in terms of Bert Similarity. 
    \item Action or Edge Similarity Threshold is used to reject connected edges or walks in the evidence graph during Edge Matching. We set this value to 0.5, making sure only edges or walks that match beyond this threshold with their action field in terms of Bert Similarity. Edge threshold is only applied to connected edges or walks. 
    \item Visual Similarity Threshold is set to 0.9 for all types of image features (sematic, face, place, object, scene-text, caption). If any 3 of these pass the threshold, we consider the image matched visually. 
    \item Edge Support Threshold is the minimum fraction of supported edges for Graph match; this is set to 0.3. 
    \item Graph Conflict  Threshold is set to the number of conflicts we tolerate. We don't tolerate any conflict, and this threshold is set at zero. 
\end{itemize}
\subsection{Cautious use of Large Language Model (LLM)}
We are careful not to use LLMs to make the final decision about the veracity of a claim. Our use of LLMs is restricted to generating ER graphs and Search terms. We also don't use LLM to build any dataset or synthetic data to train models. We leverage the NLP abilities of LLM to detect entities and relationships. While we have experimented with Mistral, Orca, and llava, we found GPT-3.5\-turbo from Openai to be the most useful in terms of the quality of the generated ER graphs and following our instructions regarding graph structure and output format. We use GPT-3.5\-turbo for all our LLM tasks.

\section{Results}
The goal of this work was to fact-check social media posts using relevant evidence from news articles. This involved retrieving evidence and verifying it through comparisons with the claim. Our framework's effectiveness depends on retrieving relevant evidence to make meaningful verification. In previous sections, we discussed evidence retrieval achieved through human-in-the-loop processes and defined our comparison framework. In the following section, we present the results of automated verification by comparing claims with the retrieved evidence.
\begin{table}
\setcounter{magicrownumbers}{0} 
\small
 \caption{ Zero shot Verification Accuracy. Note how Human-in-the-loop (Hil) leads to better accuracy.    }
 \label{tab:dataset_verf_acc}
 \setlength\tabcolsep{2 pt} 
  \begin{tabular}{|l|l|l|l|l|l|l|l|l|l|l|l|l|l|}\hline
     &    & & &  & \% TP &\% FP  & \% TN & \% FN   \\ 
    Acronym   & \#N  & \#0  & \#1  &  \% ACC  & (1-1) & (0-1)  & (0-0)  & (1-0)    \\ \hline
    B             & 154   &103   &51   & 70.77   &  88.23      &   37.86     &  62.13       &  11.76  \\
    B (HiL)        & 154   &103   &51   & 75.97   &  92.45      &   32.67     &  67.32       &  07.54  \\
    M1            & 42    &16    &26   & 78.57   &  73.07      &   12.50     & 87.50        &  26.92 \\
    M2            & 48    &17    &31   & 77.08   &  74.19      &   17.64     &  82.35       &  25.80 \\
    O             & 20    &11    &9    & 70.00   &  88.88      &   45.45     &  54.54       &  11.11 \\ \hline
    N             & 7233 &3616  &3617  & 86.21   &  95.76      &   23.34     &  76.65       &  04.23  \\ \hline
  \end{tabular}
 \end{table}

\subsection{Zero shot Verification (Graph Match and Image Match)}  
We present the primary verification results in Tab.\ref{tab:dataset_verf_acc}. Given that the Fake samples are the true class, we believe that False Negatives (predicting Fake news as Pristine) is worse than False Positives (predicting Pristine news as Fake). 

The factual political content of dataset `B', which  mostly talks about specific people at specific locations or dates, implies that we are able to disambiguate fakes easily indicated by the lower False Negative rates. However for the partitions involving posts about immigrants we often find real facts mixed with hateful prejudice or bias. This grain of truth in the fake claim leads to higher False Negatives. While the graph can clearly point out the unverified or falsified parts of the claim in the output,  the binary decision can only be corrected with a higher threshold. For our experiments we set the same thresholds for all partitions. Uniform thresholds across diverse datasets may not be optimal. Customizing thresholds based on dataset characteristics can lead to better accuracy and reliability.

Our second takeaway is the improvements in verification accuracy due to better evidence collected through the \textbf{Human-in-the-Loop (HiL)} approach.
When incorporating the Human-in-the-Loop (HiL) approach, accuracy improves to 75.97\%, and the False Negative rate drops to 7.54\%. This demonstrates the significant impact of human intervention in  the retrieval of good quality evidence. The reduction in False Positives from 37.86\% to 32.67\% further underscores the benefit of HiL in improving detection precision.

\subsection{Comparison with State-of-the art}  
\begin{table}[t]
\setcounter{magicrownumbers}{0} 
\small
 \caption{ Comparison with State-of-the-art: Our Primary observation is our competitive results without any \textit{supervision}. While our system fails to verify some real news, it does better than others in rejecting fake news. \textit{`Knw'} refers to knowledge or evidence and  \textit{`Sup'} refers to supervision.}
 \label{tab:ooc1}
 \setlength\tabcolsep{3 pt} 
 \begin{tabular}{|l|l|l|l|l|l|l|l|l|l|}\hline
            &  Method                                 & Knw                &  Sup                 &  \multicolumn{3}{|c|}{NewsCLIPpings}             &  \multicolumn{3}{|c|}{Remiss} \\ \hline  
            &                                         &                    &                      &  \multicolumn{3}{|c|}{Accuracy}             &  \multicolumn{3}{|c|}{Accuracy}\\ \hline  
            &                                         &                    &                      &  Overall           &  Fake      & Pristine  & Overall           &  Fake       & Pristine\\ \hline
 \rownumber &  CLIP\cite{luo-etal-2021-newsclippings} &                   &   $\checkmark$       &  66.1              &   56.4         &  75.7         &                   &            &           \\ 
 \rownumber &  CCN\cite{abdelnabi2022open}            & $\checkmark$       &   $\checkmark$       &  84.7              & 84.8       & 84.5      &                  &           &         \\ 
 \rownumber &  RED\cite{papadopoulos2023red}          & $\checkmark$       &   $\checkmark$        &  87.9              &            &           &                  &            &          \\ 
 \rownumber &  VTA                                    & $\checkmark$        &  $\checkmark$        &  87.4              & 86.4        & 88.4       &  42.7          & 39.5         & 46.3         \\  \hline
 \rownumber &  TS                                     & $\checkmark$        &                   &  74.5              & 82.3        & 66.5       &  57.85           & 78.5          & 32.1         \\  
 \rownumber &  RAV                                    & $\checkmark$        &                   &  86.21             & 95.76       & 76.65      &  76.15         & 83.03          & 70.28        \\   \hline

 \end{tabular}
 \end{table}
Our results on the binary task of Disinformation Detection are presented in Tab. \ref{tab:ooc1}. The related methods and baselines used are
\begin{itemize}
    \item CLIP \cite{luo-etal-2021-newsclippings} does not use any evidence but passes image and text through separate encoders to learn a binary Classification task. 
    \item CCN \cite{abdelnabi2022open} proposes the use of Cross Evidences to learn a binary Classification task. 
    \item RED \cite{papadopoulos2023red} highlights the need to point to relevant evidence. They create a dataset of relevant irrelevant evidence based on cosine similarity with the claim and train a binary classification task of fake or not that leverages this idea of relevance. As noted earlier, this allows them to point to the evidence that led to the decision, but they can not process the evidence in a fine-grained manner to say which parts of the evidence led to the decision. 
    \item Baseline \textbf{TS} is our similarity-based baseline, where we find thresholds from the validation set of NewsCLIPings. For the visual elements,  the similarity is similar to RAV, where we consider the cosine distance of pretrained features about semantics, place, and objects. For text, we use cosine distance between Bert embeddings. 

    \item Baseline \textbf{VTA} is our baseline is a supervised transformer-based setup. It is similar to the CCN Method in terms of the features used, but instead of using a memory network to capture relevant evidence,  it uses an end-to-end transformer framework trained on the final label. 
    
\end{itemize}
Our results validate that  RAV is comparable with state-of-the-art methods while maintaining high accuracy in rejecting fakes despite the zero-shot setting. 
The inclusion of evidence leads to better results, as can be inferred from the improvements due to  CCN over CLIP,  validating the idea that fact-checking should be evidence-based. But in general, the supervised evidence-based fact-checking models perform similarly on the task of detecting fakes versus pristine; our baseline VTA model performs comparable to the State-of-the-art but often deems claims as Pristine even without credible evidence or any evidence at all. The learning, however, does not transfer well to the Spanish Fake News, which can be attributed to the quality and style of Remiss data. 

\begin{figure}
\begin{center}
\includegraphics[width=0.5\textwidth]{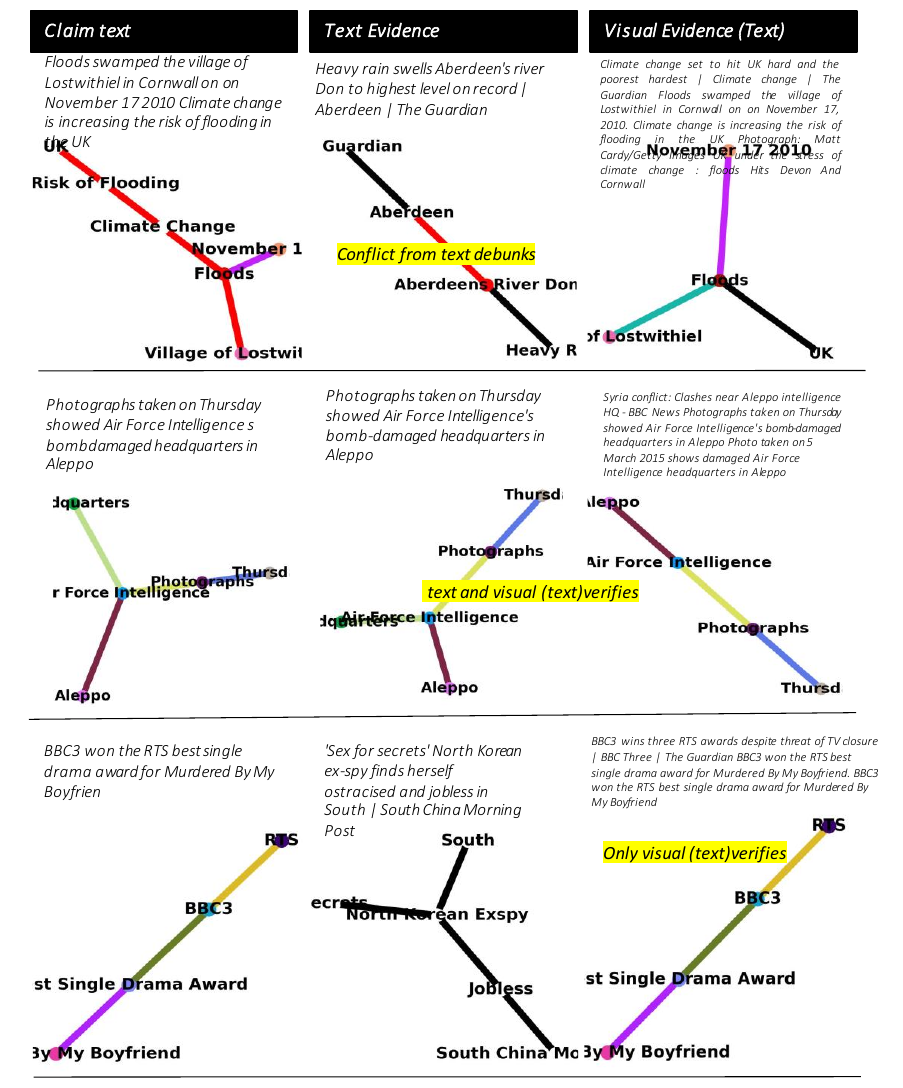} 
\caption{We present examples of Entity Relationship (ER) graphs across Claim Text (T), Text Evidence (XT), and context text from Visual Evidence (XVT). Matching nodes and edges (or walks) are color-coded, while conflicts in location or date data within node neighborhoods are marked in red, along with their edges. In the first example, a conflict arises due to location discrepancies—specifically, `Village of Lostwithiel' versus `Aberdeen.' In the second example, both pieces of evidence align with the claim, showing no conflicts. In the third, though Text Evidence (XT) does not share entities or relationships with the claim, no location or date conflict exists. Here, the verification relies on XVT, which proves useful only when the visual evidence (XV) corresponds with the claim's content (V).  }\label{fig:grapher}
\end{center}
\end{figure}

We believe that detecting fakes is more important than verifying pristine. This can be easily achieved with high similarity thresholds, as we can see in our baseline TS. Our improvements over baseline TS highlight the discriminative power of our ER graph representation over global word embeddings, given that both the methods use the same visual channel. As shown in  Fig.\ref{fig:grapher}, representing the text as Entity Relationship enables us to highlight details relevant to our task. The first example shows that while both the Text Claim and the Text Cross Evidence are about \textit{Floods in the UK}; structured representation identifies a conflict between the location  \textit{`Aberdeen} and \textit{ `Village of Lostwithiel'}. Because we only match entities against other instances of the same entity type, and not complete sentences, we can set high similarity thresholds for nodes and avoid false positives. In this specific case, both the villages are from the UK, which might mean their semantic embeddings are similar, leading to a False Positive match. We deal with this through our \textbf{hierarchical representation} and exact matching scheme as discussed in  Sec.\ref{sec:build_graph}.For locations and Dates  we use a hierarchical representation - thus, in this example, `Village of Lostwithiel' is actually represented in the graph as $ent_{type}: LOCATION, data: Lostwithiel, Cornwall, UK$ and `Aberdeen' as $ent_{type}: LOCATION, data: Aberdeen, unk, Scotland$.  

For Remiss, the texts are very different, and global semantics are less effective. High thresholds help reject most evidence when using the baseline TS, but it is only when we introduce fine-grained structured representation through RAV that we are able to identify \textbf{supports} and \textbf{conflicts}.

\begin{table*}
\vspace{-10pt} 
\small
\begin{center}
\caption{Role of Components: Graph Match, Image Match. Between XT and XV the stronger Channel is marked by Green and the weaker by Blue.  Human-in-the-loop (HiL) leads to stronger performance from retrieved Visual evidences }\label{tab:ROC}
\begin{adjustbox}{width=\textwidth}
\begin{tabular}{|l|r r r r r r|r r r r r r|r r r r r r|r r r r r r|r r r r r r|r r r r r r|}\hline
 IM    (V-XV)                 & \multicolumn{6}{|c|}{$\checkmark$}    &  \multicolumn{6}{|c|}{$\times$}    &  \multicolumn{6}{|c|}{$\times$}        &  \multicolumn{6}{|c|}{$\times$}      &  \multicolumn{6}{|c|}{$\checkmark$} \\
 GM    (T-XVT)                &  \multicolumn{6}{|c|}{$\times$}       & \multicolumn{6}{|c|}{$\checkmark$}     &  \multicolumn{6}{|c|}{$\times$}        &  \multicolumn{6}{|c|}{$\times$}      & \multicolumn{6}{|c|}{$\checkmark$} \\\hline
 Sim   (T-XT)                 &  \multicolumn{6}{|c|}{$\times$}        &  \multicolumn{6}{|c|}{$\times$}     &  \multicolumn{6}{|c|}{$\checkmark$}    &  \multicolumn{6}{|c|}{$\times$}      &  \multicolumn{6}{|c|}{$\checkmark$} \\
 GM    (T-XT)                 &  \multicolumn{6}{|c|}{$\times$}        &  \multicolumn{6}{|c|}{$\times$}     &  \multicolumn{6}{|c|}{$\times$}        &  \multicolumn{6}{|c|}{$\checkmark$}  &  \multicolumn{6}{|c|}{$\checkmark$} \\ \hline
 DS                           &N  &B &B(HiL) & M1 &M2 &O                               &N  &B &B(HiL) & M1 &M2 &O                         &N  &B &B(HiL) & M1 &M2 &O                          &N  &B &B(HiL)& M1 &M2 &O                                  &N  &B &B(HiL)& M1 &M2 &O  \\   \hline

 Acc                           & \cellcolor{blue!35}74.54   &52.59  & \cellcolor{green!35}57.79  & \cellcolor{green!35}69.04  &  \cellcolor{green!35} 72.91 & 45.00        & 70.92 & \cellcolor{green!35}55.84 &57.14  & 57.14 & 58.33 & \cellcolor{green!35}65.00     & 60.99 &\cellcolor{blue!35}46.10  & \cellcolor{blue!35}46.10 & 57.14 & 52.08 & 60.00      & \cellcolor{green!35}74.82  &43.50  &43.50 & \cellcolor{blue!35}  64.28 & \cellcolor{blue!35} 72.91 & \cellcolor{blue!35} 55.00      & 86.21  &70.77  &75.97  & 78.57  & 77.08   & 70.00\\       
 TP (1-1)                     & 99.99   &90.19  & 92.45  & 92.30  &  100.0 & 100.0        & 98.78 &72.54 &73.58  & 65.38 & 61.92 & 88.88     & 58.58 &98.11 & 98.11 & 46.15 & 35.48 & 66.66       & 97.01  &100.0  &100.0 & 73.07 & 87.09 & 66.66      & 95.76  &88.23  &92.45   & 73.07  & 74.19   & 88.88\\   
 FP (0-1)                     & 50.85   &66.01  & 60.39  & 68.75  &  76.47 & 100.0        & 56.94 &52.42 &51.48  & 56.25 & 47.05 & 54.54     & 36.58 &81.18 & 81.18 & 25.00 & 17.64 & 45.45       & 47.37  &86.13  &86.13 & 50.00 & 52.94 &  54.54     & 23.34  &37.86  &32.67   & 12.50  & 17.64   & 45.45\\    
 TN (0-0)                     & 49.14   &33.98  & 39.60  & 31.25  &  23.52 & 00.00        & 43.05 &47.57 &48.51  & 43.75 & 52.94 & 45.45     & 63.41 &18.81 & 18.81 & 75.00 & 82.35 & 54.54       & 52.62  &13.86  &13.86 & 50.00 & 47.05 &  45.45     & 76.65  &62.13  & 67.32  & 87.50  & 82.35   & 54.54\\     
 FN (1-0)                     & 00.05   &09.84  & 07.54  & 07.69  &  00.00 & 00.00        & 01.21 &27.45 &26.41  & 34.61 & 38.70 & 11.11     & 41.41 &01.86 & 01.86 & 53.84 & 64.51 & 33.33       & 02.98  &00.00  &00.00 & 26.92 & 12.90 &  33.33     & 04.23  & 11.76  & 07.54  & 26.92   & 25.80   & 11.11\\ \hline

 \end{tabular}
\end{adjustbox}
\end{center}
\end{table*}
\subsection{Ablation studies: Role of components}  In Tab. \ref{tab:ROC}  we compare the roles of various components across datasets, focusing on the contributions of Graph Match (GM) and Image Match (IM) to overall verification accuracy. The table distinguishes between textual and visual channels, marking the stronger channel with green and the weaker one with blue. Our primary observation  are as follows :

\paragraph{  \textbf{Visual Evidence Dominance }}  Visual Evidence (XV) consistently outperforms Textual Evidence (XT) in most datasets. This indicates that images play a crucial role in the verification process, likely due to their ability to provide concrete and verifiable details that are harder to manipulate compared to text. For 5 of the 6 partitions of Remiss dataset, stronger performance came from Visual Evidence (XV), obtained through feedback-based retrieval. This is more pronounced for the datasets which had missing  reverse search (XT) evidence like in the bcn19 (B) dataset.

\paragraph{  \textbf{  Human-in-the-Loop}} Human controlled search approach  leads to improvements in B(HiL) over B along both the visual channels showcasing the value of human feedback in refining the retrieval.

\paragraph{  \textbf{Graph Match}  is better than \textbf{Text Similarity} in identifying supports and conflicts }
Sim Match (T-XT) component, focuses on semantic similarity matching between textual claims and evidence  and is highlighted as essential in certain configurations. However, in most cases we can improve upon it with our fine-grained Graph Matching approach GM(T-XT). Graph Match (GM) The Graph Match component, in the T-XVT and T-XT configurations, shows varying degrees of impact across datasets. In datasets where XT retrieval is less effective (such as missing reverse search results), Graph Match T-XVT becomes more critical. However, when combined with visual evidence, its relative importance can diminish as visual cues provide more direct verification, emphasizing the need for multi-modal approaches.

\section{Analysis} 
In the following section we present some of our qualitative results and analyze them, focusing on differences between lab generated datasets and real world fake news. 

\subsection*{Nature of Text Claims}
For NewsCLIPpings, the claim texts are sourced from News websites, thus have a particular journalistic format characterized by objectivity,  relevant details, and brevity; in Remiss, the texts are sourced from social media posts where there is a subjective retelling of the claims, often with a strong bias. This verbosity affects the  visual evidence \textbf{XV retrieval } and  the \textbf{T-XT comparison}. 

\subsection*{Nature of Visual Claims}
The NewsCLIPpings images are mostly professionally taken pictures already published on news websites. For Remiss, the images are mostly taken by individuals and often differ in visual perspective from the ones reported by journalists on news websites, as seen in Fig. \ref{fig:qua00} and Fig.\ref{fig:quamiss}. While the top left sample is pristine in both the figures and  the corresponding visual evidences retrieved are also correct, it is the widely varying perspective between the claim and evidence in  Fig.\ref{fig:quamiss} lead to Out-of-Context. The unpublished nature of Social media visual claims affect the Text Evidence \textbf{XT retrieval}, while their varied perspective affects the \textbf{V-XV comparison}

\subsubsection*{ \textbf{XV Retrieval}}
We use direct search with the claim text T to retrieve Visual Cross evidence XV aiming to verify the claim image V. Because of the publisher origin of NewsCLIPpings, the style of the text also acts as a clue and often returns exact text matches and thus exact visual evidence matches, as seen in Fig. \ref{fig:zs_newsclip_v}, row 1. For Remiss, we never encounter exact text matches from direct search, and spend the bulk of our effort in the retrieval stage, with a Human in the loop \textbf{feedback-based retrieval} resulting in visual evidence matches, eg, Fig.\ref{fig:qua00}. Our feedback retrieval is particularly useful when it comes to finding visual evidence matches for verbose texts claims that are subjective retelling of some existing story, benefiting from an objectively structured text representation. In Fig. \ref{fig:fdr_remiss} we see its application to the real world remiss dataset, where we often have to go to multiple search term refinement to find the correct image.
Even for the returned search results, which are mostly from Google's cache, the link cannot often be traced back to its source for contextual text. While we did encounter cases of page update or broken links as a cause, the major reason was actually paywalls. Finally, retrieval is also limited by its actual coverage of the claim story in the news websites. 

\subsubsection*{ \textbf{XT Retrieval}} For NewsCLIPpings, the reverse search with image claim to retrieve XT is successful for around \textbf{70\%} of images samples. This is despite the already published nature of the source images. For Remiss, reverse search with an image rarely returns exact matches. In all we find XT evidence for only \textbf{17\%} of the total image samples processed, as detailed in Tab \ref{tab:dataset_split}.  We hypothesize that the absence of published exact forms in mainstream media is one of the causes.

\subsubsection*{\textbf{ V-XV comparison}}
Visual similarity, from a verification perspective, is a challenge on its own. Even with the good quality visual evidences of NewsCLIPings, courtesy already published T, the IM(V-XV) comparison is mostly doing well (99.99\%) in Fake detection Task, ie TP(1-1),  and not verification (49\%),  as can be seen in Tab. \ref{tab:ROC}.  In fact use of contextual text GM(T-XVT) leads to a drop in performance hinting that the performance is mostly due to exact matched images and high visual similarity thresholds. For Remiss, the visual evidence XV is often from varied perspective and its only when we use the contextual text we are able to able improve upon the verification accuracy at the cost of False Negatives. Implying we mostly find images from alternate visual perspectives with similar contextual text. While contextual text extracted from visual evidence can lead to more interpretable support or conflict, our results illustrate that it degrades performance in all cases, and should only be used conditional on high visual similarity.

\subsubsection*{\textbf{ T-XT comparison}} 
\begin{figure}
\begin{center}
\includegraphics[width=0.45\textwidth]{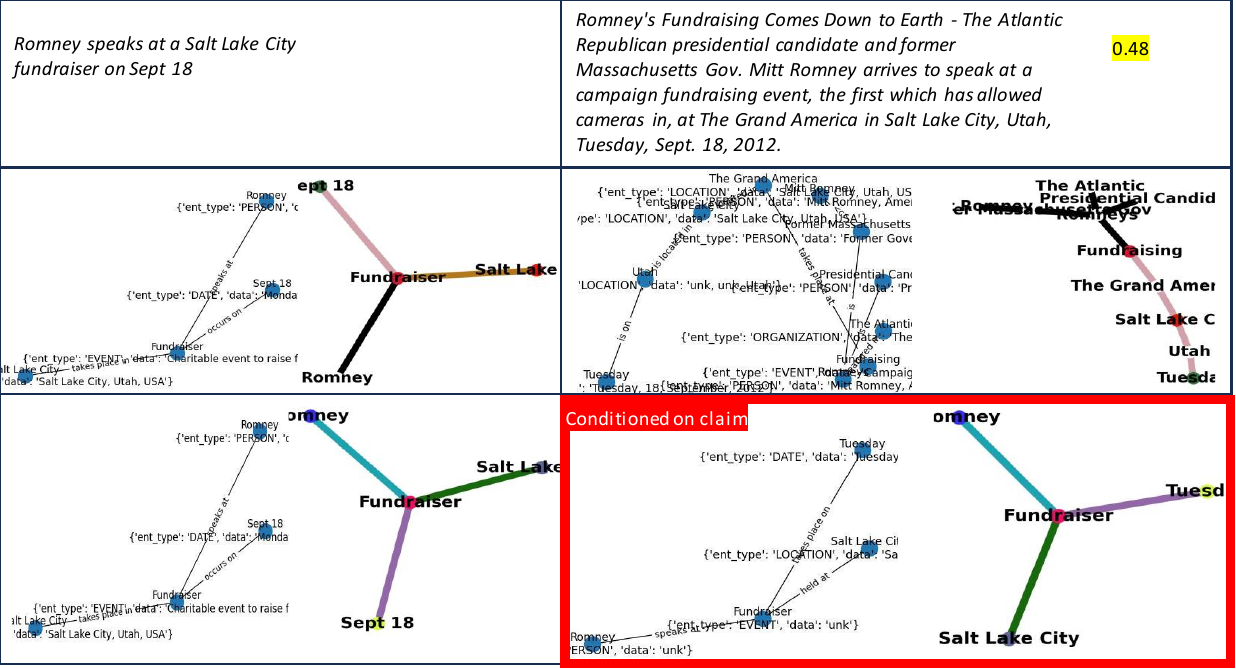} 
\caption{Using a Conditional Graph helps preserve structure. In the top row, we compare two texts with a low BERT similarity score of 0.48. Despite low similarity, common nodes and edges are detected, shown in matching colors in the simplified annotated graph to the right. In the bottom row, conditioning entity and relation detection on the claim generates structurally aligned graphs, enabling detection of additional nodes and edges. }\label{fig:congra}
\end{center}
\end{figure}

Text Evidence, when available from exact matches, is telling. NewsCLIPpings text claims originate as the scraped captions of image claims, which are usually short and concise, but when searched with image claim the retrieved text evidence can often be more detailed. This difference is style and structure of the claim text and evidence text is more pronounced in Real World fake news samples from Remiss. This necessitates \textbf{Conditional Graph building} as discussed in Sec. \ref{bcond_gra}, where the evidence graph construction is focused on entities and relationships shared with the claim, as seen in Fig \ref{fig:congra}. 
The conditional graph is the principal reason why we see improvement in overall accuracy for NewsCLIPings. While the fine grained graph structured matching leads to improved False Negatives across datasets it can not compensate for the lack of relevant good quality Text evidence of Remiss. 

\begin{figure}
\begin{center}
\includegraphics[width=0.49\textwidth]{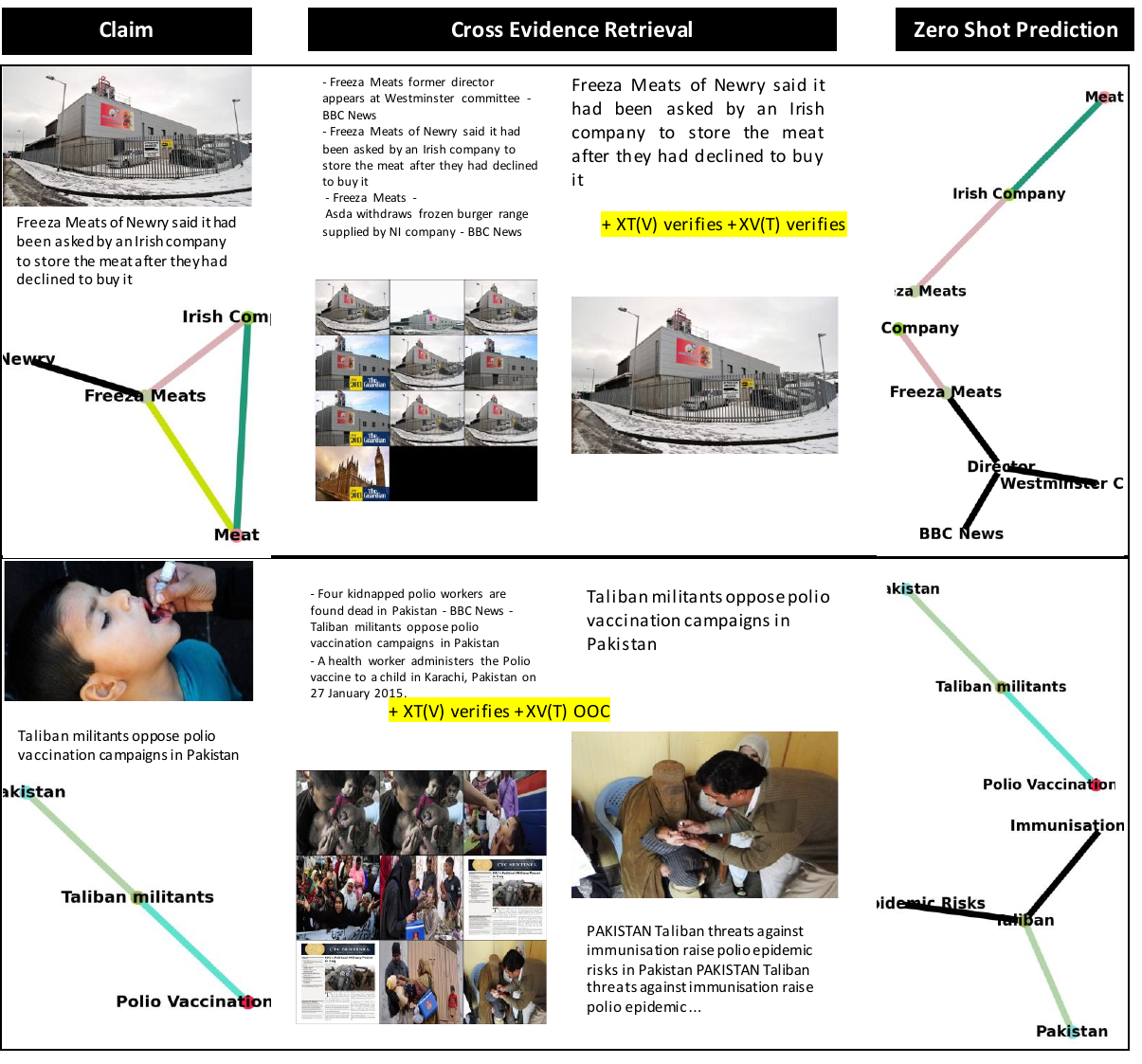} 
\caption{Verification in NewsCLIPpings }\label{fig:zs_newsclip_v}
\end{center}
\end{figure}

\begin{figure}
\begin{center}
\includegraphics[width=0.49\textwidth]{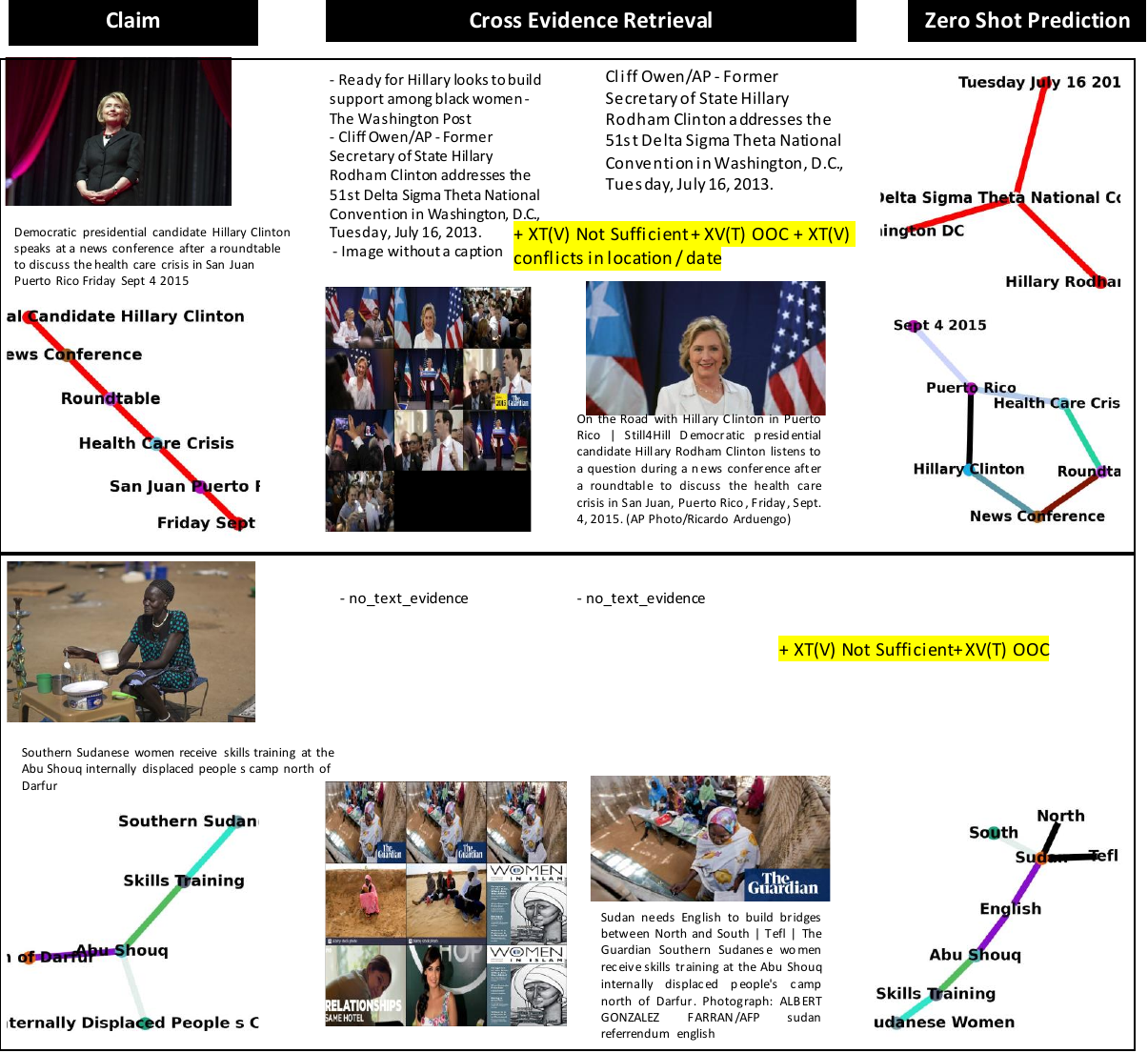} 
\caption{Fake Detection in NewsCLIPpings }\label{fig:zs_newsclip_f}
\end{center}
\end{figure}

\subsection{Success Cases}

\begin{figure}
\begin{center}
\includegraphics[width=.49\textwidth]{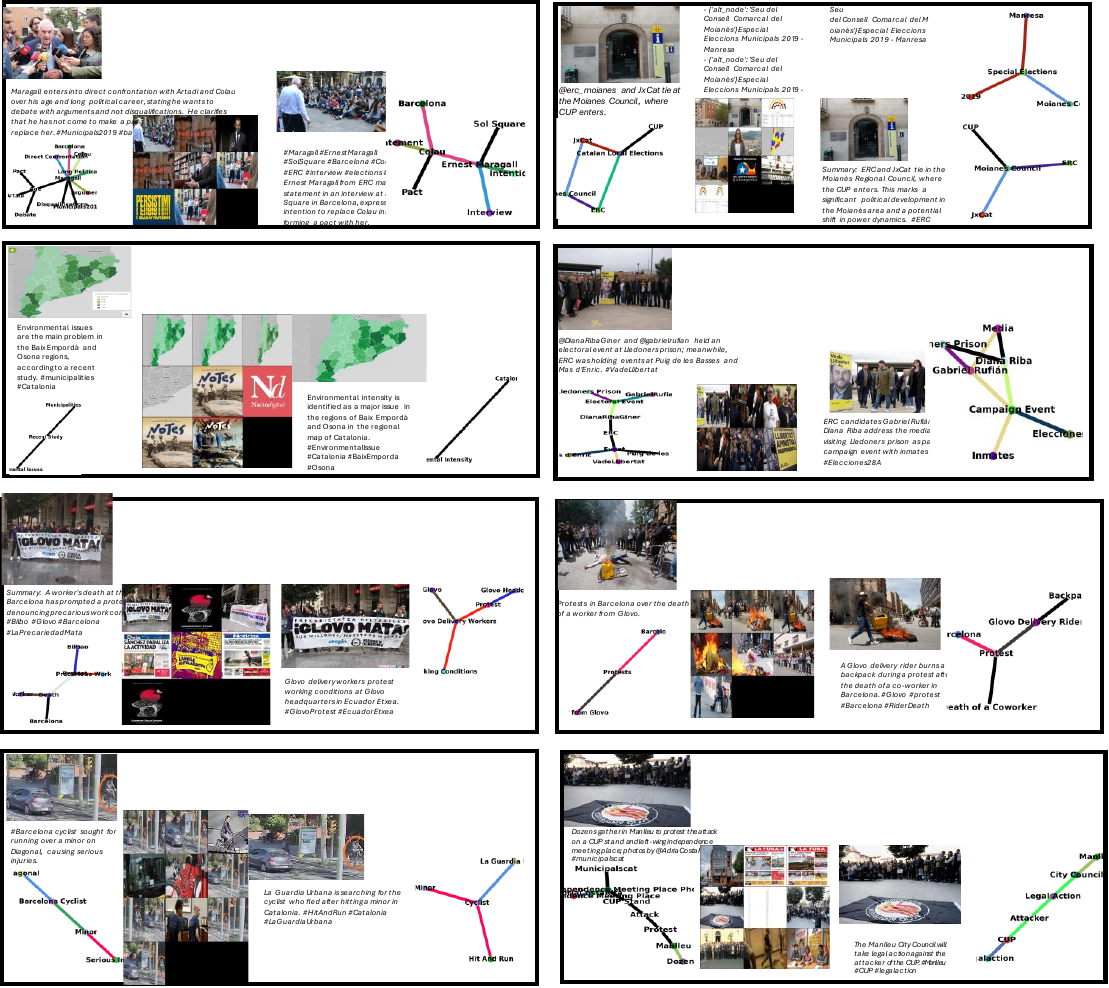}
\caption{Verification in Remiss}\label{fig:qua00}
\end{center}
\end{figure}

\paragraph{\textbf{Detecting Pristine}}
In Fig. \ref{fig:zs_newsclip_v} and Fig. \ref{fig:qua00}  we see examples of Pristine News from NewsCLIPings and Remiss respectively that has been successfully verified by external news sources. While for NewsCLIPings we were able to find exact matches, for Remiss  the reverse search with image mostly fails, and it is the visual evidence retrieved by querying with text claim that leads to verification. We see these are mostly reports about protests, gathering or crime, which are usually well covered in news websites. Also note that while we have no explicit way to deal with claims that contain maps or charts, but when retrieved as an evidence the visual feature similarity itself can verify the claim in case of exact matches. 

\begin{figure}
\begin{center}
\includegraphics[width=.49\textwidth]{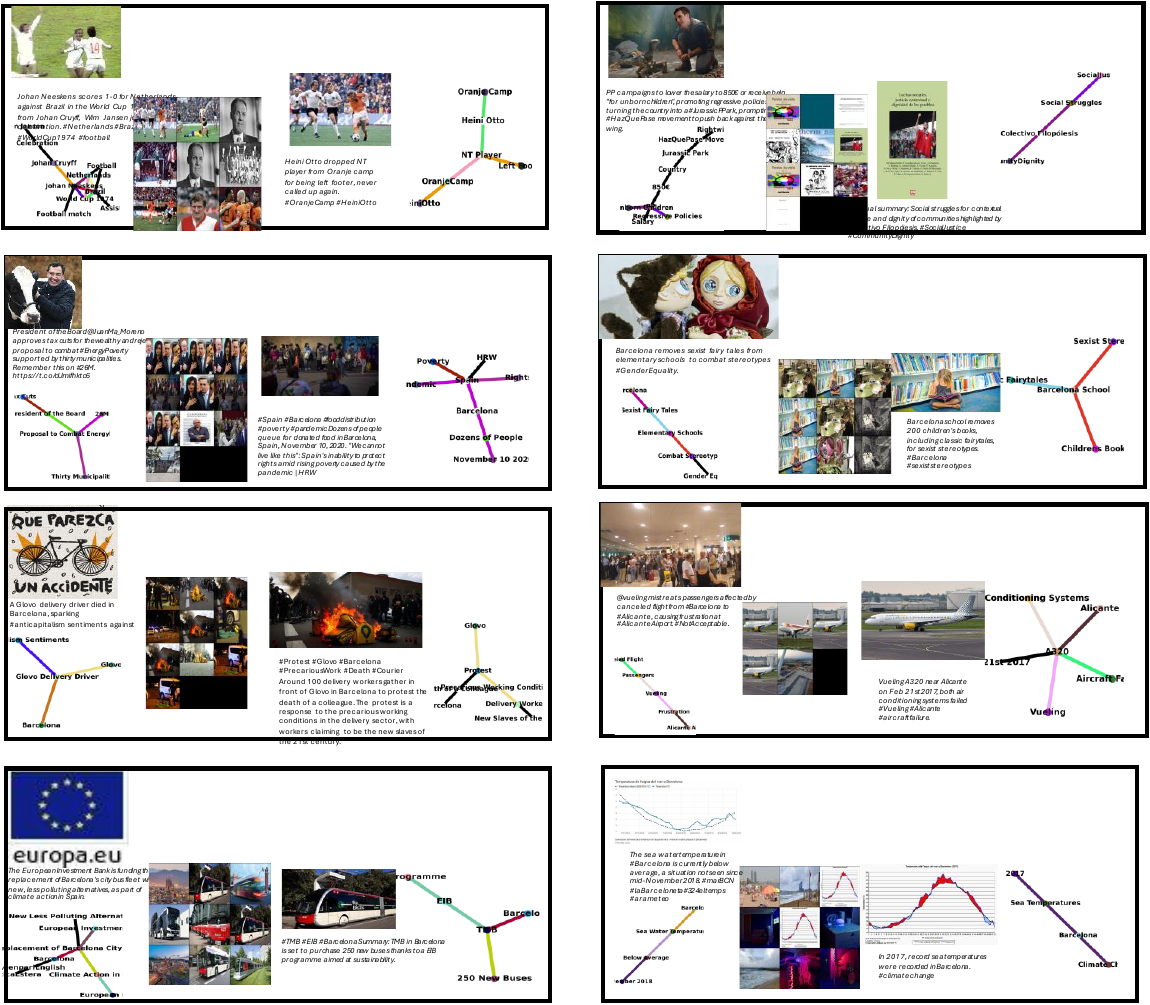}
\caption{Fake Detection in Remiss. With Top left as 1,1, we see that (2,1),(2,2),(3,1), and (4,1) are Out-Of-Context }\label{fig:qua11}
\end{center}
\end{figure}

\paragraph{\textbf{Detecting Fakes}}
In Fig. \ref{fig:zs_newsclip_f} and Fig. \ref{fig:qua11}  we see examples of Fake News from NewsCLIPings and Remiss respectively that we successfully reject. The `fake' label here follows the definition we introduced in Sec. \ref{whatisf}, that is based on evidence present. Apart from these samples with poor evidence marked as fake, we also have the Out-of-Context pairs marked. Note that despite the similarity in terms of claim text and visual evidence text XVT, the samples are marked as fake because of lack of Visual Similarities. In particular we look at the images corresponding to a food delivery partner, we see that in Fig .\ref{fig:qua00} (3,2) when the text is paired with relevant contextual image in the claim we mark it as Pristine, but when the text is paired with a symbolic image, in  Fig. \ref{fig:qua11} (3,1), we mark the sample as Out-of-Context.


\subsection{Fail Cases}
\begin{figure}
\begin{center}
\includegraphics[width=.48\textwidth]{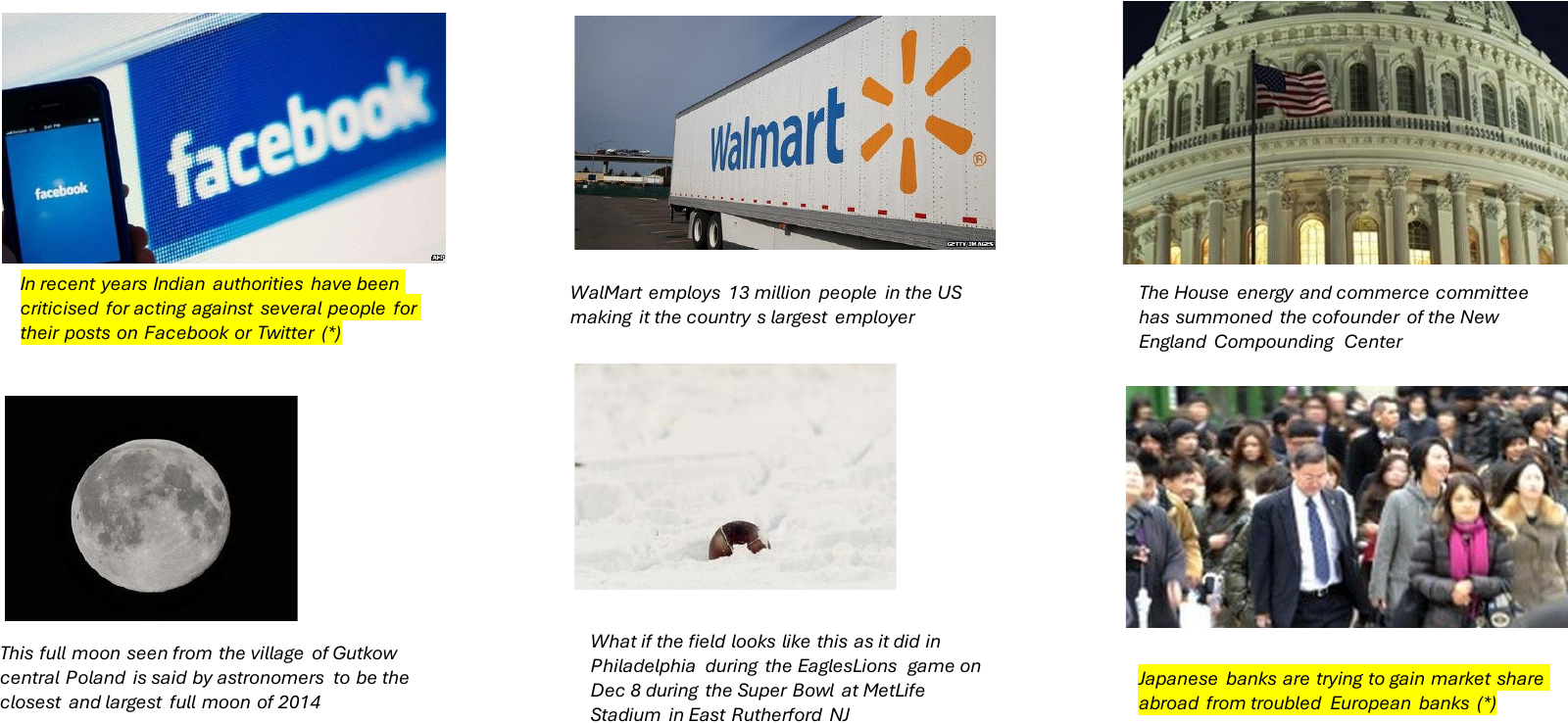}
\caption{  NewsCLIPpings Fail Cases: We Mark pristine samples as fake, because the visual claims are symbolic, note how the VTA baseline often marks these as pristine, marked with yellow}\label{fig:ccnd_sym}
\end{center}
\end{figure}

\begin{figure}
\begin{center}
\includegraphics[width=.48\textwidth]{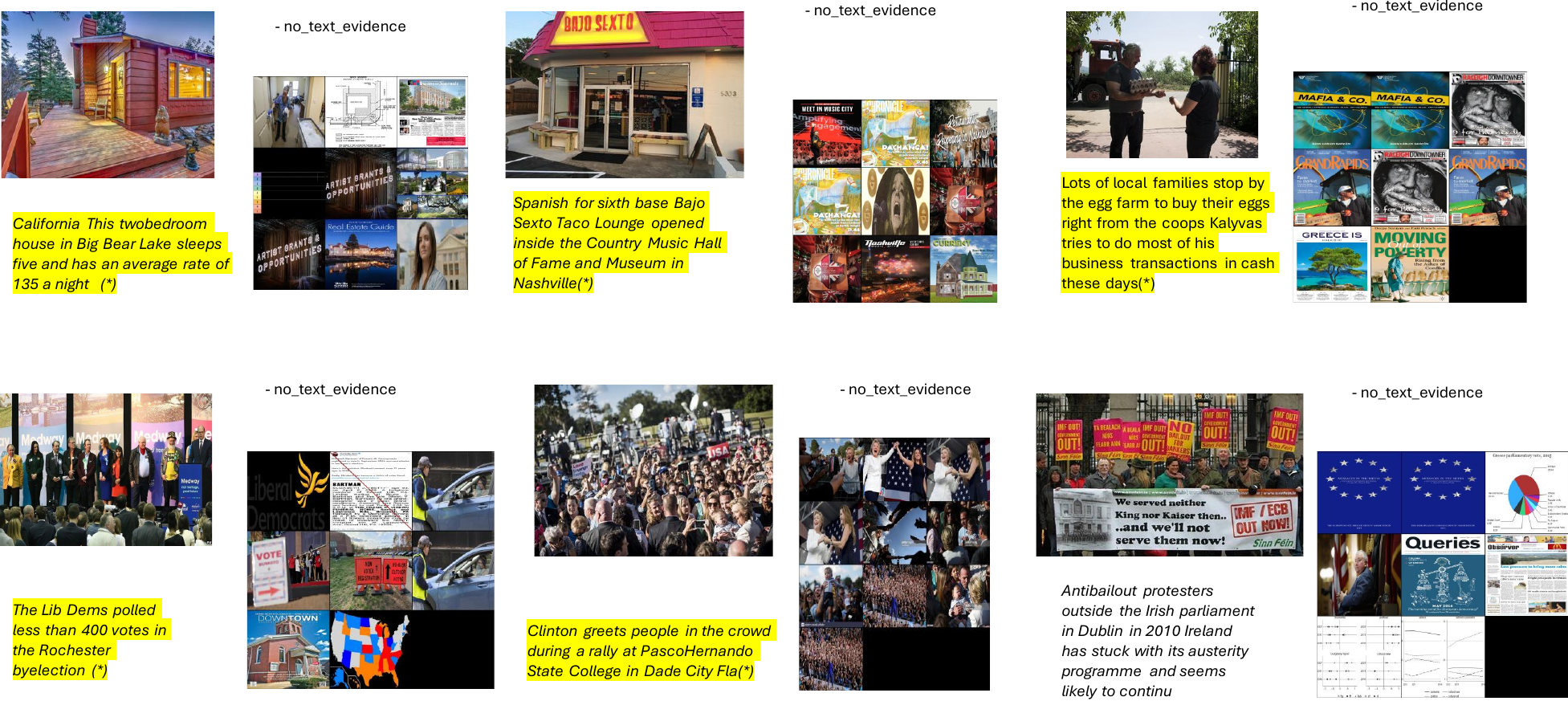}
\caption{ NewsCLIPpings Fail Cases: We Mark pristine samples as fake, because the visual claims are could be not found, note how the VTA baseline often marks these as pristine despite the lack of evidence, marked with yellow}\label{fig:ccnd_bad}
\end{center}
\end{figure}

In Fig. \ref{fig:ccnd_sym} and \ref{fig:ccnd_bad} we highlight some of our fail cases in NewsCLIPings. 
In Fig. \ref{fig:ccnd_sym} we reject samples as fake, because the visual content is symbolic, and thus the  visual claim can not be meaningfully compared with the visual evidence from the text. In Fig. \ref{fig:ccnd_bad}  we show how often even with poor evidence the trained VTA Model makes predictions which are correct. Our RAV based model rejects all these as fakes, due to the lack of evidence. 

\begin{figure}
\begin{center}
\includegraphics[width=.49\textwidth]{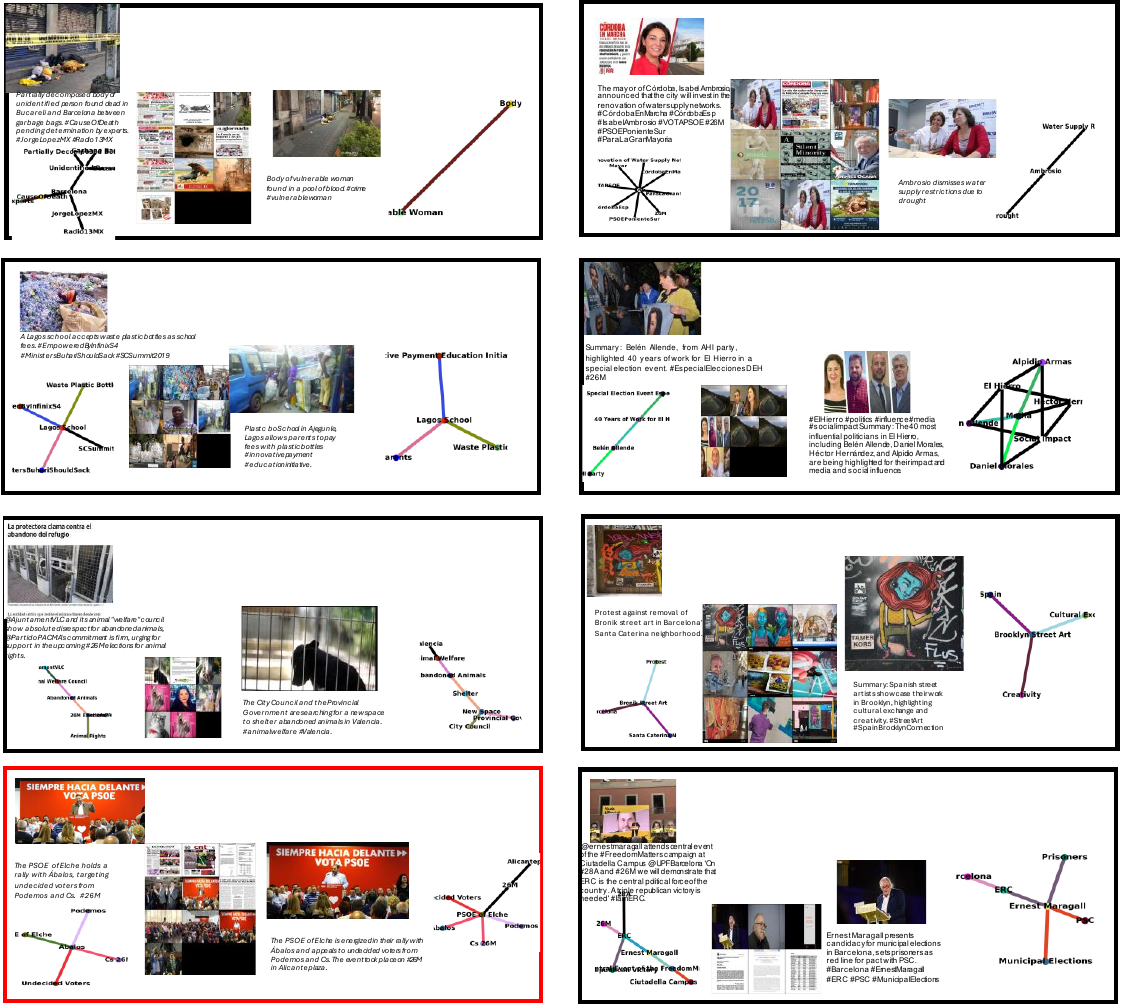}
\caption{Remiss Fail cases. Samples in the left show examples of Pristine news which could not be verified, and on the  right column we see fake examples verified by incorrect evidence.}\label{fig:quamiss}
\end{center}
\end{figure}
In Fig. \ref{fig:quamiss}  we see examples where our framework failed for Remiss. Our failure to verify pristine samples  could be attributed to the lack of similar visual evidence. The left column corresponds to these claims for which the retrieved visual evidence was not similar enough. Exception is the sample in the last row(4,1), highlighted in red, where an incorrect visual evidence was used to make the correct prediction that this is a pristine news. The image chosen as evidence belongs to the same event but has a different speaker but with caption that aligns with the claim text.

As discussed earlier, we had marked samples with inconclusive evidence as fake. In Fig. \ref{fig:quamiss} right column  we see such examples which, despite inconclusive evidence, was marked pristine.  For the top two samples which are about political personalities, we did find related evidence which mentions them but it was judged by a human annotator as not being enough to verify it and thus marked as fake. 
For the third example, we see that a similar street art by the same artist verifies the claim, but we can clearly see that while the character may seem the same they  differ in posture and are in different physical locations. For the last example, while both the claim and evidence feature the same person giving speeches in similar settings, they are in different locations. This location data however was only present for the claim and thus it could not be used to find conflict with the evidence, and visual similarity and semantic relatedness marked the sample as pristine.
 
\subsection{Contrasting Our Verification Tool with Fact-Checking Sites and Search Engine Results}

We have developed a web-based verification tool for multimodal social media posts, allowing users to upload both text and images, configure settings, retrieve external evidence, refine searches, and visualize the verification results through support and conflict analysis between claims and evidence. An example of this tool's functionality is shown in Fig. \ref{fig:demo_example}.

\begin{figure}
\begin{center}
\includegraphics[width=.54\textwidth]{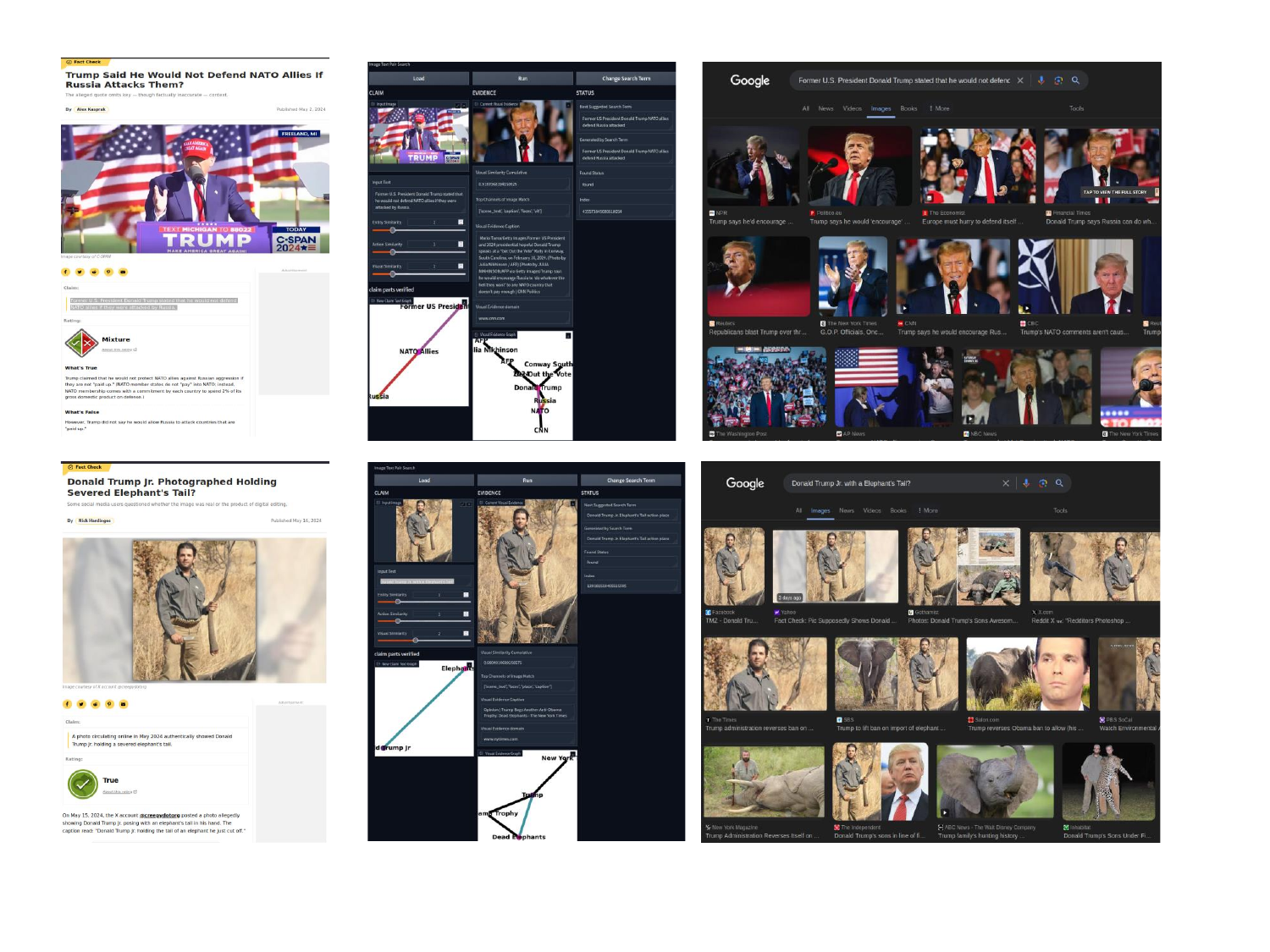}
\caption{Successful Verification.}\label{fig:ver_eng}
\end{center}
\end{figure}

\begin{figure}
\begin{center}
\includegraphics[width=.49\textwidth]{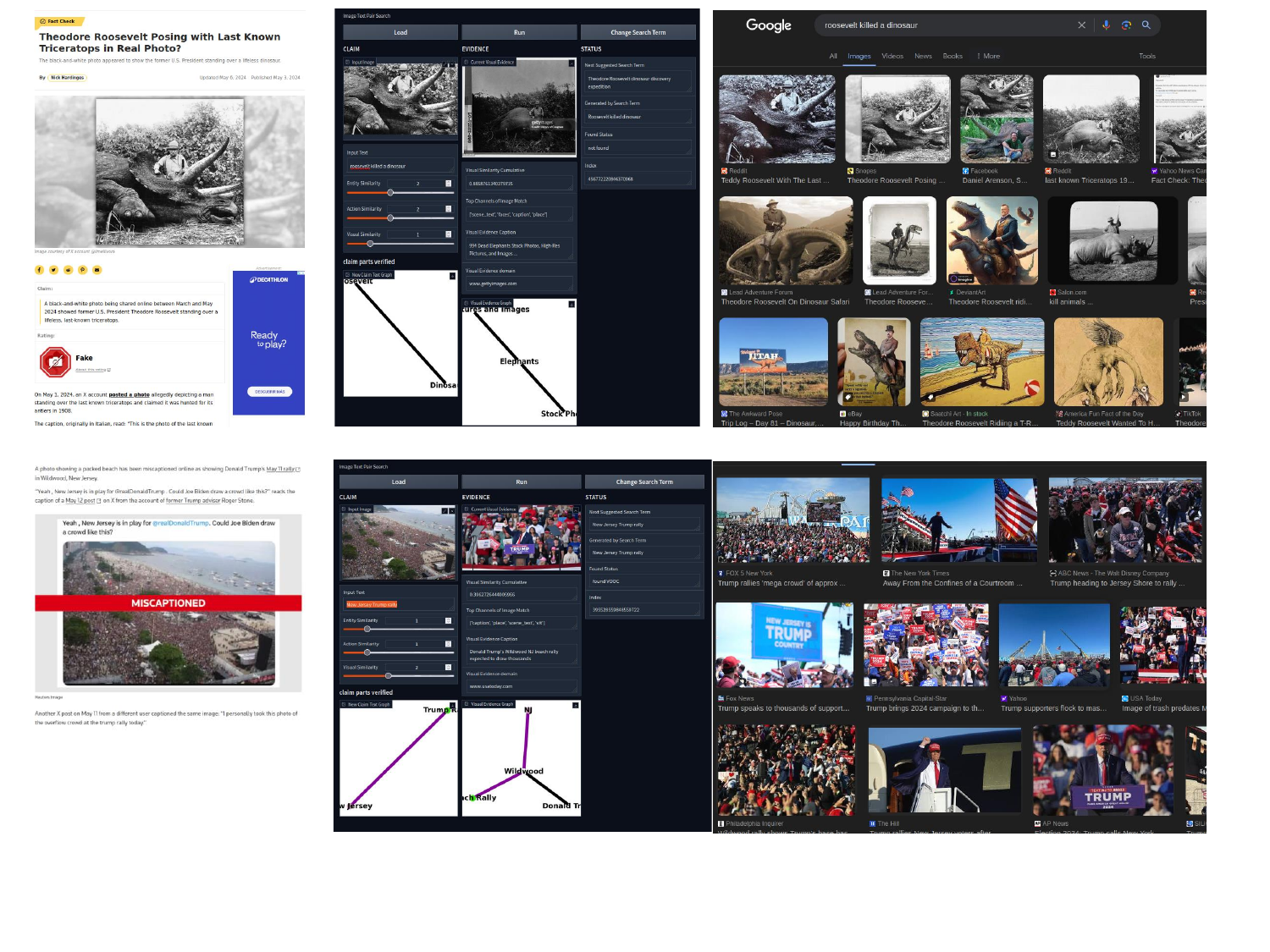}
\caption{Successful Rejection.}\label{fig:rej_eng}
\end{center}
\end{figure}

\begin{figure}
\begin{center}
\includegraphics[width=.49\textwidth]{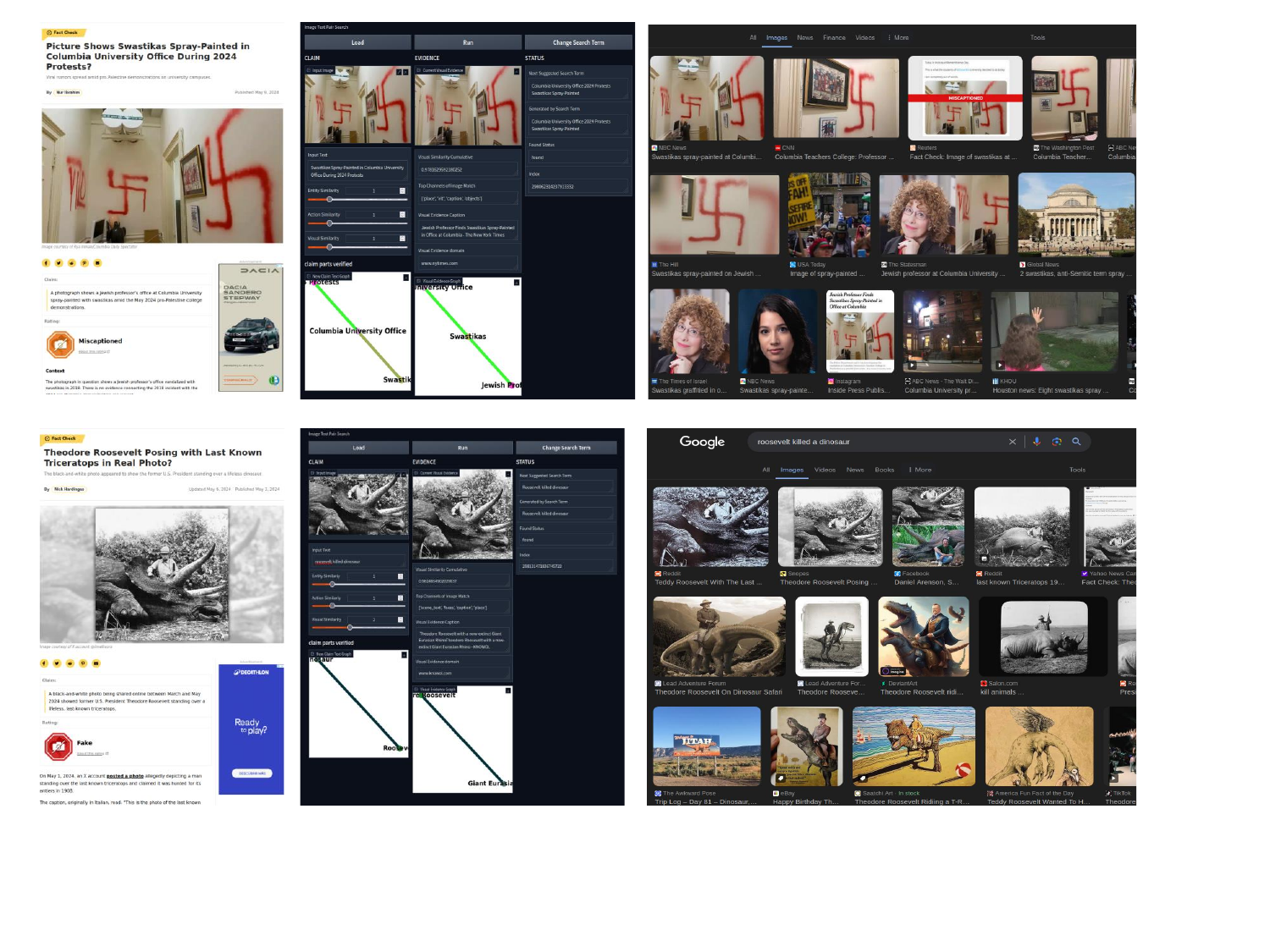}
\caption{Fail Cases: In the First case, the claim tries to re-purpose an image from 2018. But the retrieved evidence does not include the date `2018', and this lack of date detail led to the claim being verified. In the second case we show what if do not restrict ourselves to credible news sources, then malicious sources can validate our claims }\label{fig:miss_eng}
\end{center}
\end{figure}

In  Figs. \labelcref{fig:ver_eng,fig:rej_eng,fig:miss_eng},  we compare our results against Fact-Checking websites and Google Searches. 
Fact-checking sites offer  detailed explanations provided by expert Journalists. Such explanations are usually based on multiple evidence sources linking them. Google Search on the other hand provides a ranked list of hits, devoid of any explanations. Fact checking is tedious and what is true, requires human understanding. In  this work we propose a tool, that can alleviate some of the tedious elements of fact checking in regards to finding evidence. As we show in the example runs, our method is not just able to find the relevant evidences, but also highlight the supports elements in the text. 


\section{Conclusion}
In addressing the pervasive challenge of social media disinformation, this work presents a pioneering zero-shot framework for verifying multimodal claims, emphasizing clarity, interpretability, and a real-time approach to evidence retrieval. By breaking down claims into entity-relationship graphs for text and pretrained feature sets for images, we offer a structured approach that enables both in-depth analysis and transparent verification. Unlike conventional binary classifiers, our framework distinguishes itself by empowering users to visually understand which parts of the claim align or conflict with trusted evidence sources, mirroring the meticulous rigor of journalistic fact-checking.

A major advantage of our approach lies in its independence from labeled datasets and supervision, which makes it inherently free from the biases that often accompany data-driven training methods. This lack of reliance on labeled data allows our system to flexibly adapt to new and emerging events, addressing the limitations of models trained on historical data, which often struggle with recent developments.

Our analysis has highlighted several valuable insights: while high-threshold visual similarity checks effectively identify claims related to well-documented events, the retrieval process faces challenges with poorly documented or newly emerging events, as well as with highly generic or verbose claims. Despite these limitations, our system demonstrates strong potential in handling complex claims involving widely reported events like protests or high-profile gatherings. Future enhancements—such as refining visual similarity measures, enhancing metadata usage for contextual accuracy, and expanding evidence sources—promise to further increase verification accuracy and adaptability.

Ultimately, this framework advances the goal of enabling users to discern fact from fiction more confidently on social media. With an interface for human feedback and visual representation of the verification process, our system paves the way for more transparent, explainable verification tools. By providing interpretable, unbiased, and contextually accurate results, we believe this approach is a significant step towards combating misinformation in the digital age.

\section*{Acknowledgments}
The authors would like to acknowledge the Centre de Visió per Computador and Universitat Autònoma de Barcelona for their support in this research. A European patent application related to this work has been filed under the following details:

\begin{itemize}
  \item \textbf{Title:} \textit{“System and Computer-Implemented Method of Detecting Fake Multimodal Media”}
  \item \textbf{Applicants:} Centre de Visió per Computador; Universitat Autònoma de Barcelona
  \item \textbf{Application No.:} 25382241
\end{itemize}
The patent application has been submitted to the European Patent Office.

\bibliographystyle{splncs04}
\bibliography{egbib}
\newpage
\appendix
\section{Demonstration and Examples}
In this section, we present a series of images that illustrate various aspects of our work. The demonstration example showcases the primary functionality, while subsequent examples highlight specific features and use cases. Each image is carefully selected to provide clear insights and visual representation of our findings, making it easier for readers to grasp the practical applications of the presented concepts. 

\begin{figure*} 
    \centering
    \includegraphics[width=\textwidth,height=\textheight,keepaspectratio]{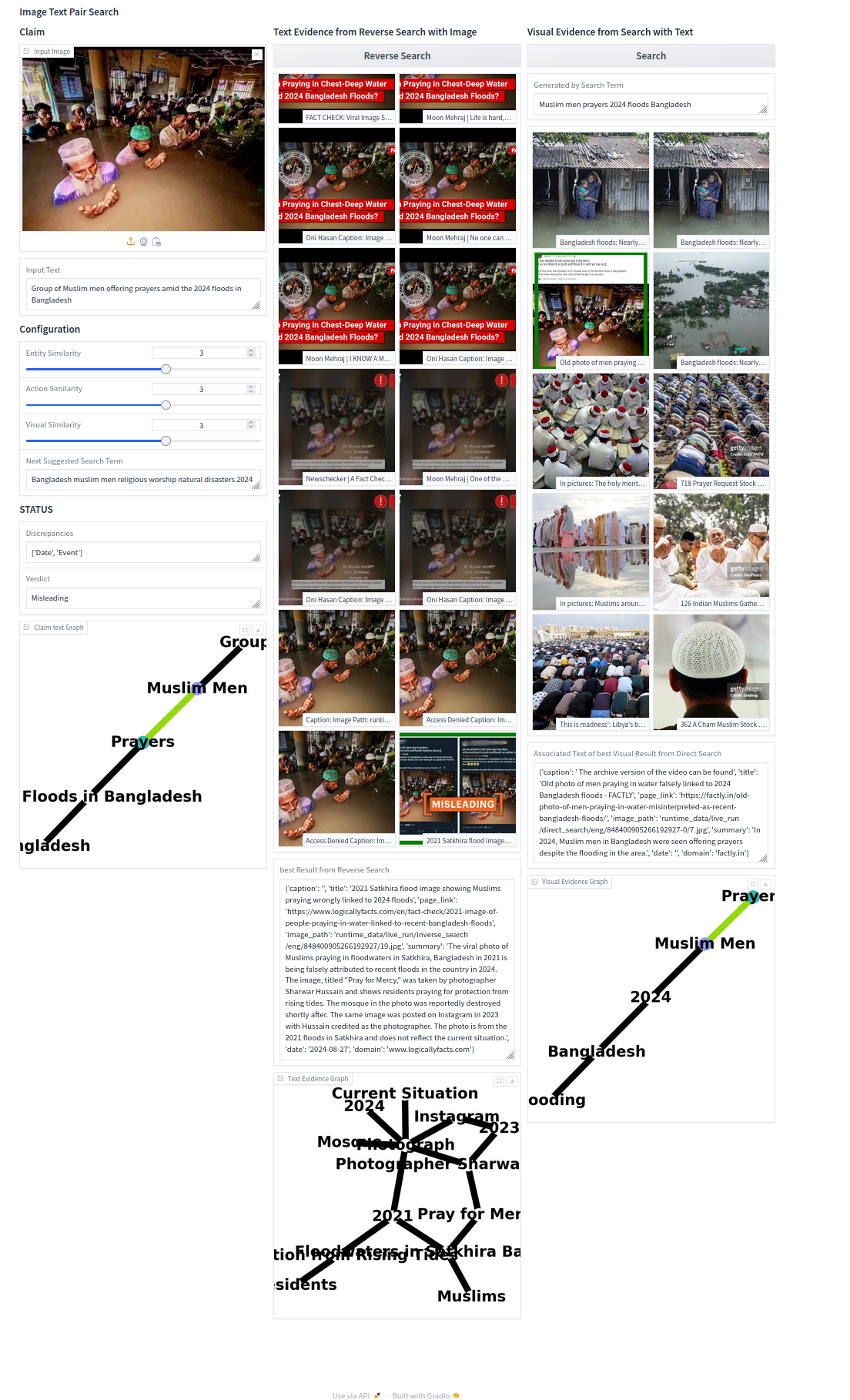}
    \caption{Example 1}
    \label{fig:ex1}
\end{figure*}
\newpage 

\begin{figure*} 
    \centering
    \includegraphics[width=\textwidth,height=\textheight,keepaspectratio]{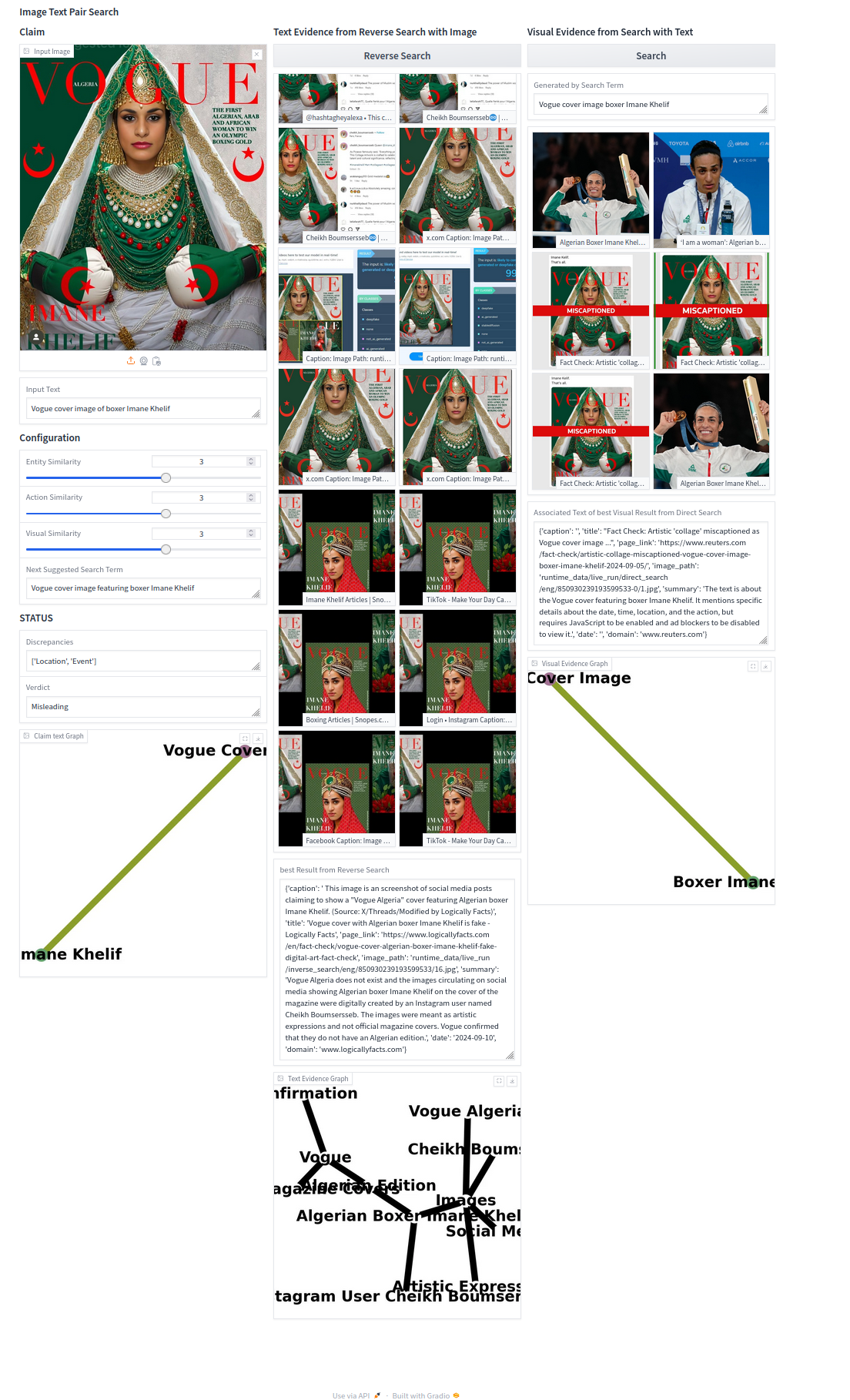}
    \caption{Example 2}
    \label{fig:ex2}
\end{figure*}
\newpage 

\begin{figure*} 
    \centering
    \includegraphics[width=\textwidth,height=\textheight,keepaspectratio]{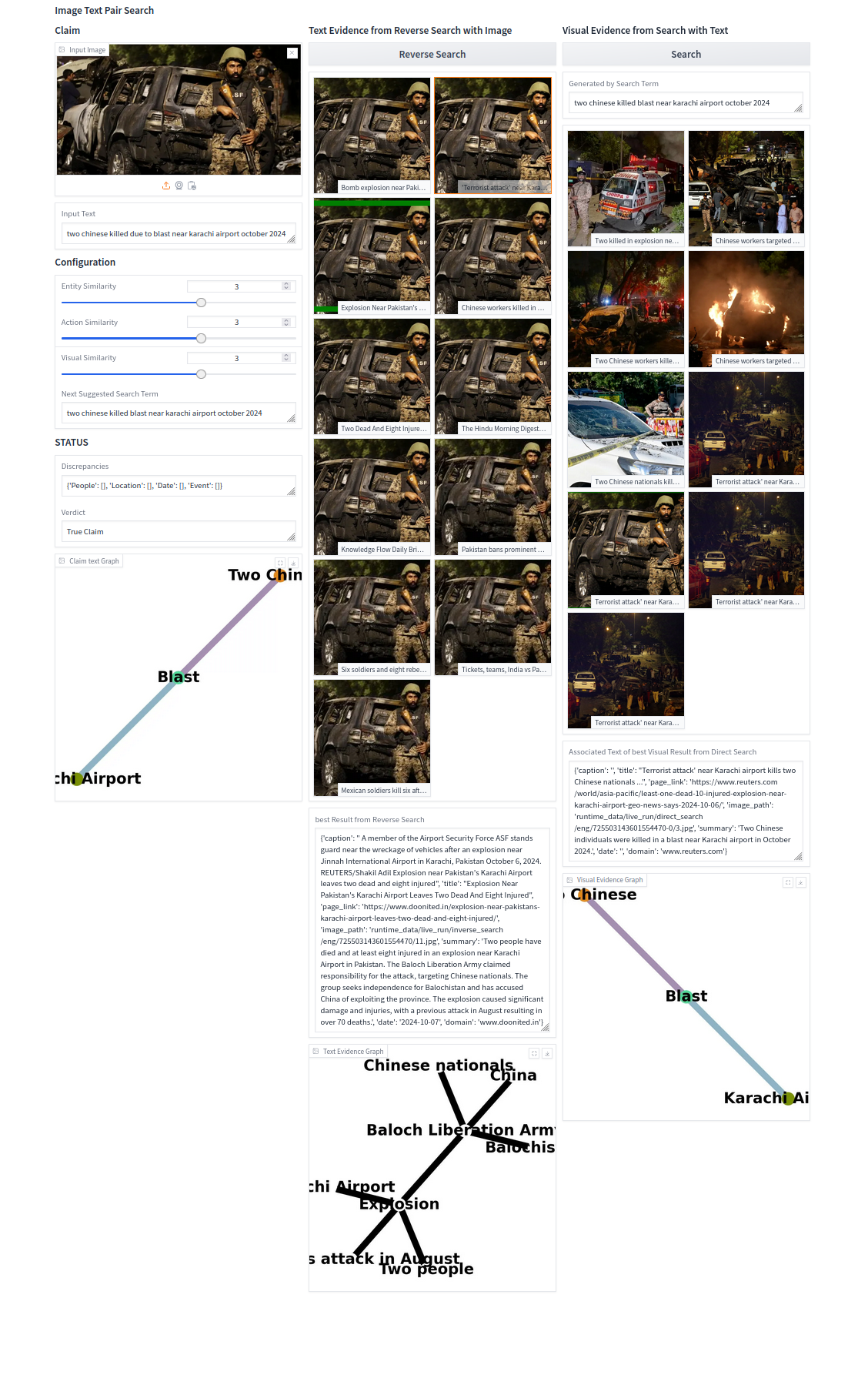}
    \caption{Example 3}
    \label{fig:ex3}
\end{figure*}

\begin{figure*} 
    \centering
    \includegraphics[width=\textwidth,height=\textheight,keepaspectratio]{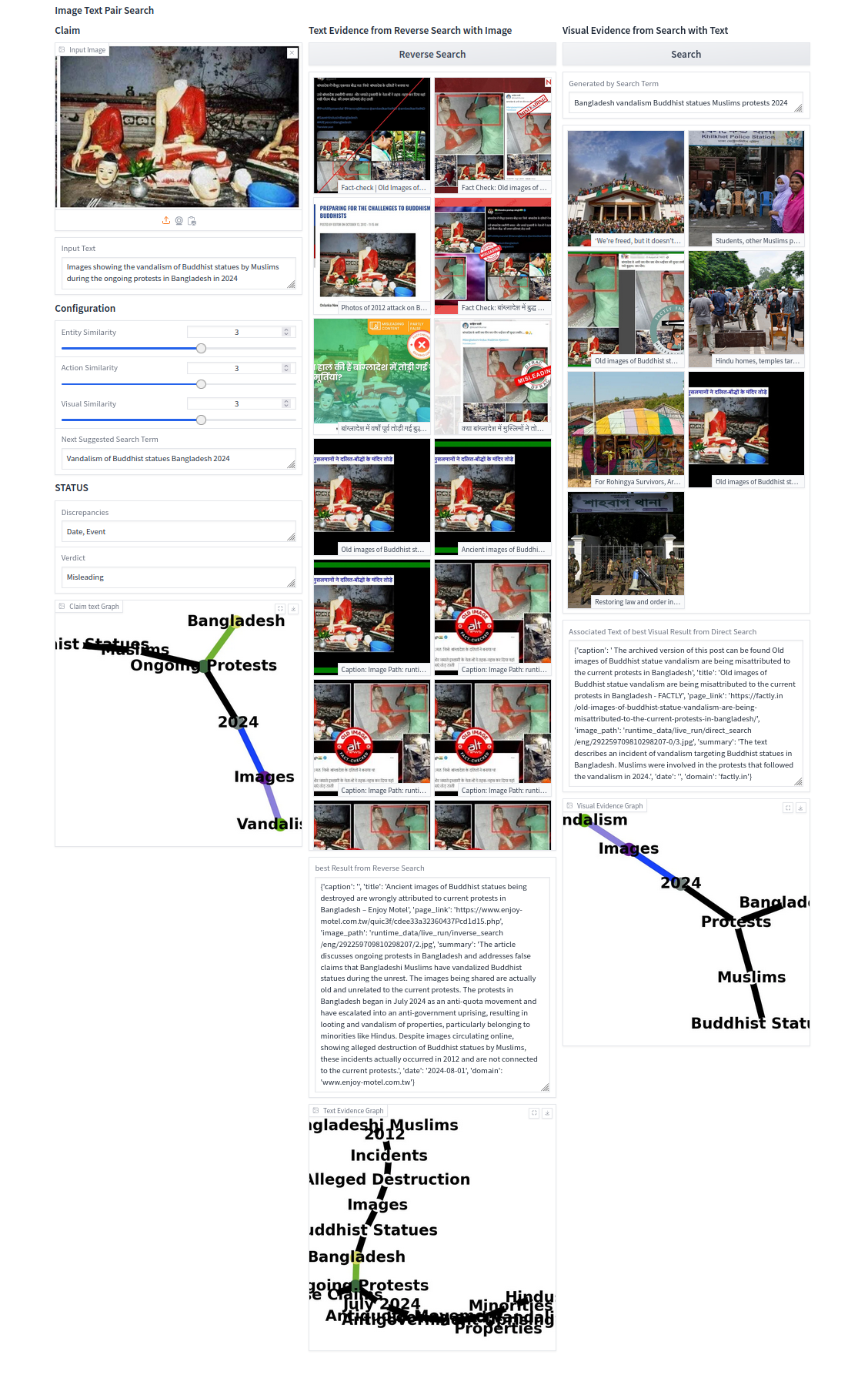}
    \caption{Example 4}
    \label{fig:ex4}
\end{figure*}
\end{document}